# Distinct topological excitonic insulators characterized by quantum geometry

Zhuowei Liu,[1] Rui Wang,[2,3,4,5,*] and Baigeng Wang[2,3,4,†]

[1]National Laboratory of Solid State Microstructures, Department of Physics, Nanjing University, Nanjing 210093, China
[2]National Laboratory of Solid State Microstructures, Nanjing University, Nanjing 210093, China
[3]Collaborative Innovation Center of Advanced Microstructures, Nanjing University, Nanjing 210093, China
[4]Jiangsu Physical Science Research Center, Nanjing University, Nanjing 210093, China
[5]Hefei National Laboratory, Hefei 230088, China

The intertwining of electron-hole correlation and nontrivial topology is known to give rise to exotic topological excitonic insulators. Here, we show that the involvement of quantum geometry can characterize more exotic excitonic phases exhibiting physical properties that are not influenced by their topology but by geometry. Starting from a topological band insulator and gradually reducing the band gap, many-body interaction can initially generate a $p + ip$-wave and then an $s$-wave excitonic insulator. Interestingly, they bear the same Chern number but exhibit completely different spin textures and magneto-optical Kerr responses, reflecting the intricate geometric distinctions in their wave functions. We also propose to enhance the correlation effect via Floquet engineering, which provides a systematic way to realize these topological excitonic insulators and their phase transitions in the nonequilibrium steady states. Our results demonstrate correlated phenomena characterized by quantum geometry, beyond the conventional topological classifications.

## I. INTRODUCTION

The interplay between correlation and topology has become one of the most active fields during the past decades. It generates a variety of correlated topological phases that significantly enhance fundamental knowledge of condensed matter physics, including fractional quantum Hall states [1–3], topological orders [4–7], fractional Chern insulators [8], etc. These correlated phases either exhibit fractionalized excitations or edge modes with topological origins, thus being robust against local perturbations.

Despite the topology, quantum geometry is another key quantity underlying generic quantum systems. In particular, the real part of the quantum geometric tensor [9] describes the quantum metric of an eigenstate space, measuring the orthogonality or the amplitude distance of quantum states under small changes [9–11]. Recent studies have revealed the important role of quantum geometry in governing many quantum phenomena in noninteracting systems, such as the Hall effect [12,13], shift currents [14–16], circular photogalvanic effect [17]. For correlated electronic systems, quantum geometry is also known to be essential [18–20]. However, its interplay with correlation is much less understood. Compared to the topology, a global property of a quantum state, the geometry contains more detailed information about the local metric. Thus, it is intriguing to ask whether there are correlated phenomena that are not characterized by their topology, but by geometry.

In this article, we show that the interplay between correlation and quantum geometry points to more exotic phases of matter beyond the topological descriptions. We reveal topological excitonic insulators (EIs) exhibiting the same bulk topology but distinct physical properties owing to their different quantum geometries. EIs are correlated insulators extensively studied in recent years [21–30]. For band insulators (semiconductors) with a gap smaller than the exciton binding energy, the electron-hole interaction is known to spontaneously induce the EI phases with $s$-wave pairing symmetry [21]. Here, we instead start from a topological insulator (TI) exhibiting spin-momentum locking [Fig. 1(a)]. Despite gradually reducing the band gap, we remarkably find that a $p + ip$ triplet EI is energetically favored in the intermediate gap regime. Only when the gap is reduced to small values does the conventional $s$-wave singlet EI become stable, as shown in Fig. 1(c). Notably, all three insulating phases, i.e., the original TI, the $p + ip$, and the $s$-wave EI, exhibit the same Chern number. However, they display completely different physical properties including the spin textures and the magnetic-optical Kerr effects. Such differences reveal the important role of quantum geometry in influencing correlated states [31–35] and their physical natures.

To observe the quantum geometry characterized topological EIs and their phase transitions, the band gap needs to be tuned. This can be achieved by Floquet engineering by applying a periodic driving field. As shown in Fig. 1, we show that a high frequency driving field can continuously reduce the quasienergy band gap. This efficiently enhances the correlation effect, stabilizing the Floquet topological EIs (FTEIs) with distinct quantum geometric properties. Our work therefore suggests (1) a systematic way to modulate the correlation effect via Floquet engineering, and (2) reveals correlated phenomena characterized by quantum geometry.

*Contact author: rwang89@nju.edu.cn
†Contact author: bgwang@nju.edu.cn

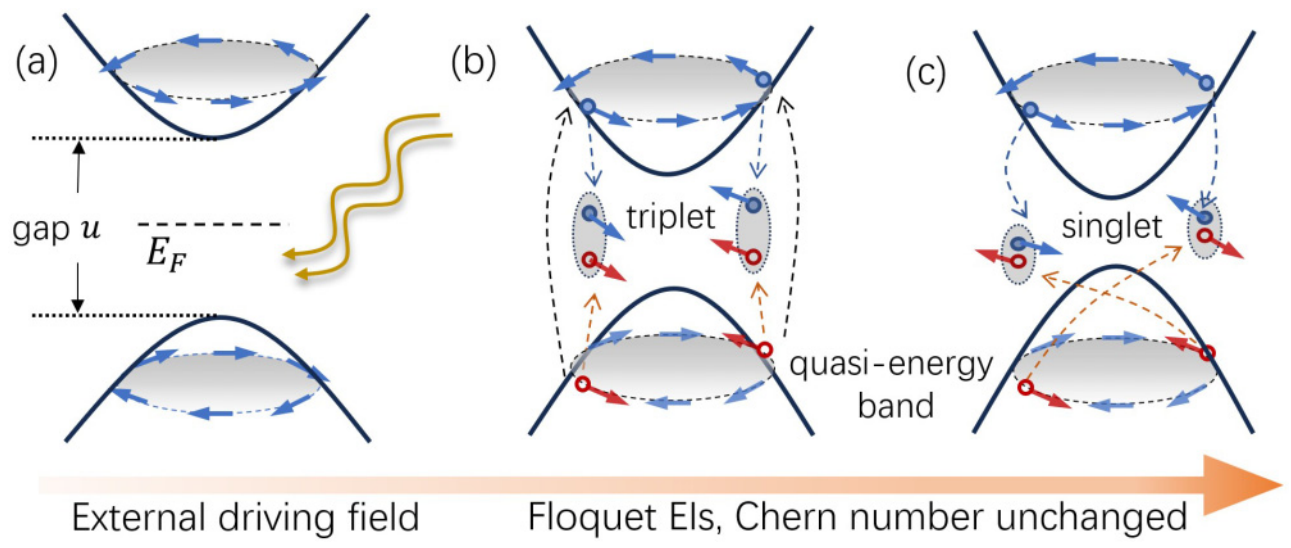

FIG. 1. (a) Starting from a TI, the external driving fields can reduce the quasienergy band gap. (b) For intermediate gap values, a $p + ip$-wave triplet EI can be favored. (c) For small gap values, an $s$-wave singlet EI is stabilized.

## II. FLOQUET-ENHANCED CORRELATION EFFECT

We consider the minimal model of an interacting topological insulator. The free part, in the basis of $\Phi_{\mathbf{k}} = [f_{\mathbf{k},\uparrow}, f_{\mathbf{k},\downarrow}]^{\mathrm{T}}$, is given by $H_0 = \sum_{\mathbf{k}} \Phi_{\mathbf{k}}^{\dagger}(v_F \boldsymbol{\sigma} \cdot \mathbf{k} + u\sigma_z)\Phi_{\mathbf{k}}$, where the Pauli matrix $\boldsymbol{\sigma} = (\sigma_x, \sigma_y)$ acts in the spin space. $H_0$ describes a Dirac fermion in two dimensions (2D) with spin-momentum locking and a gap $u$. This can be realized on the surface of 3D TIs [36], by either magnetic doping [37,38] or surface state coupling [39] in thin films consisting of several quintuple layers. An energy cutoff $v_F\Lambda$ is implicit in the low-energy effective description, and we set $v_F = 1$. On top of $H_0$, we further consider the Hubbard-type interaction, $H_{\mathrm{int}} = U\sum_{\mathbf{r}} f_{\mathbf{r},\uparrow}^{\dagger} f_{\mathbf{r},\uparrow} f_{\mathbf{r},\downarrow}^{\dagger} f_{\mathbf{r},\downarrow}$.

For large $u$, the gap dominates the interaction, resulting in a 2D TI characterized by a half-integer Chern number $C = 1/2$ [23,40,41]. We then consider a periodic in-plane electric field with frequency $\omega$ described by the vector potential $\mathbf{A}(t) = (A\sin(\omega t), A\sin(\omega t + \phi))$, where $\phi$ is the polarization angle and $A$ is the field strength. $\mathbf{A}(t)$ brings about dynamics absent in $H = H_0 + H_{\mathrm{int}}$, described by $H'(t) = v_F \sum_{\mathbf{k}} \Phi_{\mathbf{k}}^{\dagger}\boldsymbol{\sigma} \cdot \mathbf{A}(t)\Phi_{\mathbf{k}}$, leading to the total Hamiltonian $H^{\mathrm{tot}}(t) = H + H'(t)$. Using Floquet theory [42,43], the dynamics over time spans longer than the period $T = 2\pi/\omega$ can be separated from the micromotion occurring within $T$ [44]. The stroboscopic time evolution in steps of $T$ is then described by the Floquet Hamiltonian $H_F$. The latter not only generates the quasienergy but also enables tailoring of nonequilibrium steady states, a process known as Floquet engineering [45–50].

Focusing on the $l = 0$ Floquet band [44], we arrive at the second-quantized effective interacting Hamiltonian as

$$H_{\mathrm{eff}} = \sum_{\mathbf{k}} \Phi_{\mathbf{k},0}^{\dagger}(\sigma \cdot \mathbf{k} + m\sigma_z)\Phi_{\mathbf{k},0} + U\sum_{\mathbf{r}} n_{\mathbf{r},0,\uparrow} n_{\mathbf{r},0,\downarrow}, \quad (1)$$

where $\Phi_{\mathbf{k},0} = [f_{\mathbf{k},0,\uparrow}, f_{\mathbf{k},0,\downarrow}]^{\mathrm{T}}$, $n_{\mathbf{r},0,\sigma} = f_{\mathbf{r},0,\sigma}^{\dagger} f_{\mathbf{r},0,\sigma}$, and $f_{\mathbf{r},0,\sigma}^{\dagger}$ ($f_{\mathbf{r},0,\sigma}$) denotes the creation (annihilation) operator for electrons on the zeroth Floquet band. The gap $m$ is obtained as $u + \frac{A^2}{\omega}\sin(\phi)$, and we focus on the circular polarization ($\phi = -\pi/2$) below. Interestingly, $H_{\mathrm{eff}}$ has a form similar to that of the undriven model but with a renormalized quasienergy gap $m$. Since the gap is reduced, while the strength of interaction $U$ remains intact, the correlation effect is effectively enhanced. This would favor excitonic instabilities as illustrated in Fig. 1.

## III. FLOQUET TOPOLOGICAL EXCITONIC INSULATORS CHARACTERIZED BY QUANTUM GEOMETRY

We now study the Floquet excitonic phases based on Eq. (1). Excitons are bound states formed by the conduction electrons and the valence band holes. Thus, we convert to the band basis via the transformation $[c_{\mathbf{k}}, v_{\mathbf{k}}]^{\mathrm{T}} = R\Phi_{\mathbf{k},0}$, where $c_{\mathbf{k}}$ and $v_{\mathbf{k}}$ denote the annihilation operators for the Floquet modes in the conduction and valence bands, respectively. The matrix entries are $R_{11} = \cos\theta_{k,m}e^{i\phi_{\mathbf{k}}}$, $R_{12} = \sin\theta_{k,m}$, $R_{21} = \sin\theta_{k,m}e^{i\phi_{\mathbf{k}}}$, $R_{22} = -\cos\theta_{k,m}$, and $\cos\theta_{k,m} = \frac{k}{\sqrt{k^2+(\epsilon_k-m)^2}}$, $\sin\theta_{k,m} = \frac{\epsilon_k-m}{\sqrt{k^2+(\epsilon_k-m)^2}}$, where $\mathbf{k} = (k\cos\phi_{\mathbf{k}}, k\sin\phi_{\mathbf{k}})$ and $\epsilon_k = \sqrt{k^2+m^2}$. The dispersion then takes the form $E_{\pm,k} = \pm\epsilon_k$.

The second term in Eq. (1) must be accordingly projected onto the band basis. Focusing on the interband interactions in the forward scattering channel that favor excitonic instabilities [23], we arrive at the relevant interaction as [51]

$$\begin{aligned} H_{\mathrm{int}}^{\mathrm{eff}} = U\sum_{\mathbf{k},\mathbf{k}'} &[\cos^2\theta_{k,m}\cos^2\theta_{k',m}e^{i(\phi_{\mathbf{k}}-\phi_{\mathbf{k}'})} \\ &+ \sin^2\theta_{k,m}\sin^2\theta_{k',m}e^{i(\phi_{\mathbf{k}'}-\phi_{\mathbf{k}})} \\ &+ 2\cos\theta_{k,m}\cos\theta_{k',m}\sin\theta_{k,m}\sin\theta_{k',m}]c_{\mathbf{k}}^{\dagger}v_{\mathbf{k}'}^{\dagger}v_{\mathbf{k}}c_{\mathbf{k}'}. \quad (2) \end{aligned}$$

For the $\phi_{\mathbf{k}}$-independent term in Eq. (2), we introduce the order parameter $\Delta_s = U\sum_{\mathbf{k}}\cos\theta_{k,m}\sin\theta_{k,m}\langle c_{\mathbf{k}}^{\dagger}v_{\mathbf{k}}\rangle$, with isotropic $s$-wave pairing. For the $\phi_{\mathbf{k}}$-dependent terms, two order parameters can be introduced, respectively, namely, $\Delta_{p,1} = U\sum_{\mathbf{k}}\cos^2\theta_{k,m}e^{i\phi_{\mathbf{k}}}\langle c_{\mathbf{k}}^{\dagger}v_{\mathbf{k}}\rangle$ and $\Delta_{p,2} = U\sum_{\mathbf{k}}\sin^2\theta_{k,m}e^{-i\phi_{\mathbf{k}}}\langle c_{\mathbf{k}}^{\dagger}v_{\mathbf{k}}\rangle$. These describe $p \pm ip$ excitonic insulators, as they generate off-diagonal terms in the mean-field Hamiltonian that are proportional to $(k_x \pm ik_y)/k$. Inserting the order parameters into Eq. (2) and minimizing the mean-field ground state energy, we self-consistently determine the condensation energy and the phase diagram with varying $A$.

The condensation energy $E_c$, defined as the energy difference between the mean-field phases and the free state, is shown in Fig. 2(a). In our interacting system, the quantum geometry whose imaginary part is Berry curvature can be defined in the Bogoliubov basis of the mean-field Hamiltonian [52]. For small $A$ (large $m$), the band gap $m$ dominates over the interaction $U$, and the Floquet TI with $C = -1/2$ is stable. For larger $A$ with $A^2/\omega \lesssim u$, such that the gap is renormalized to small values ($m \gtrsim 0$), $U$ plays the dominant role. In this case, the excitonic order is formed with the isotropic $s$-wave pairing, i.e., $\Delta_s \neq 0$. Interestingly, in the intermediate region, we find that a $p + ip$-wave Floquet EI becomes stable, with $\Delta_{p,1} \neq 0$ and $\Delta_s = 0$. We note that the transition between different types of EIs can be explained by the relative strength of effective interaction components, as is demonstrated by

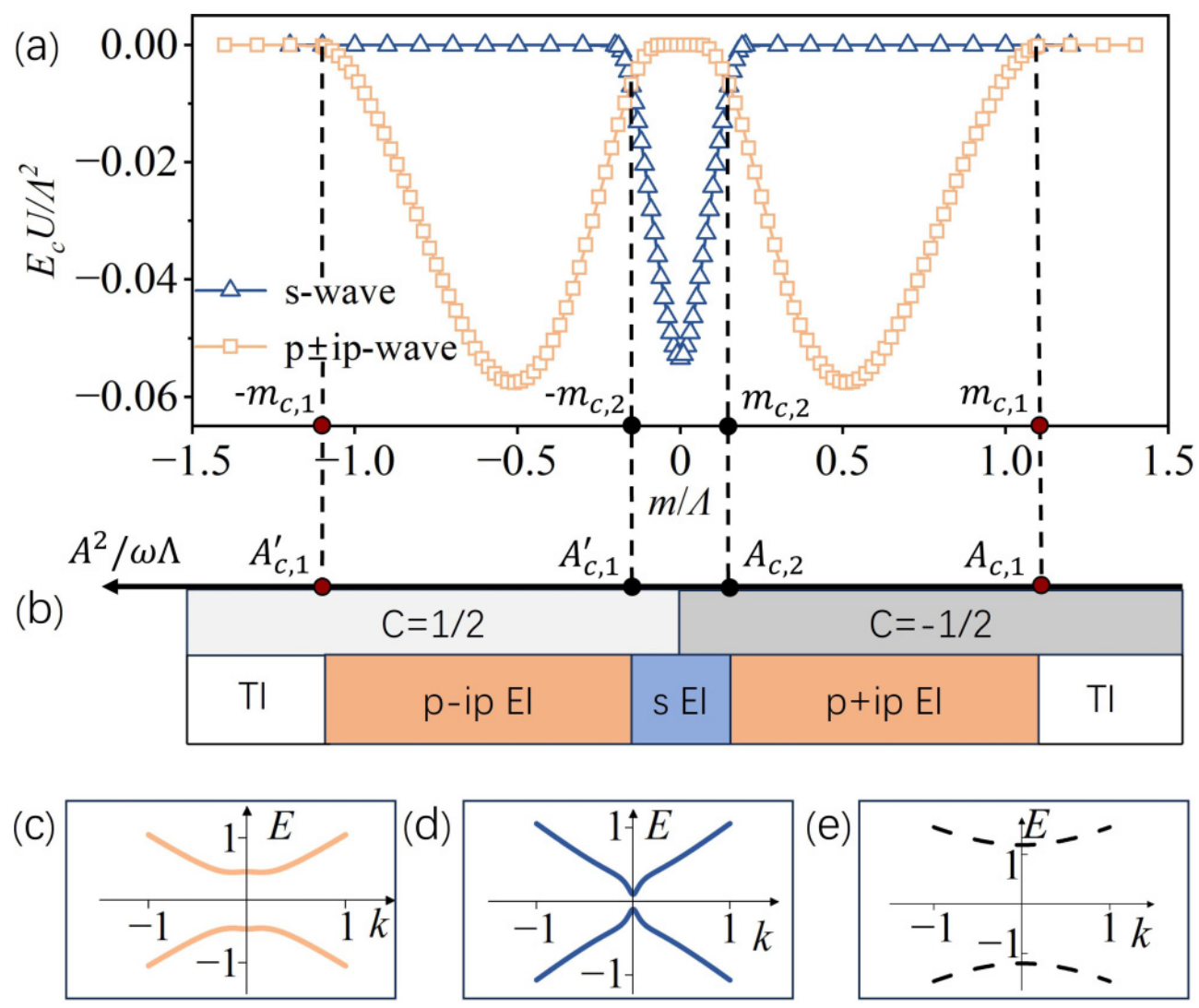

FIG. 2. (a) The condensation energy of different phases with varying the renormalized gap $m$. Two transitions occur at $|m_{c,1}| = 1.12$ and $|m_{c,2}| = 0.15$. (b) The phase diagram vs the field strength $A$. Panels (c)–(e), respectively, show the characteristic energy spectrum for the $p \pm ip$ EI, $s$-wave EI, and TI.

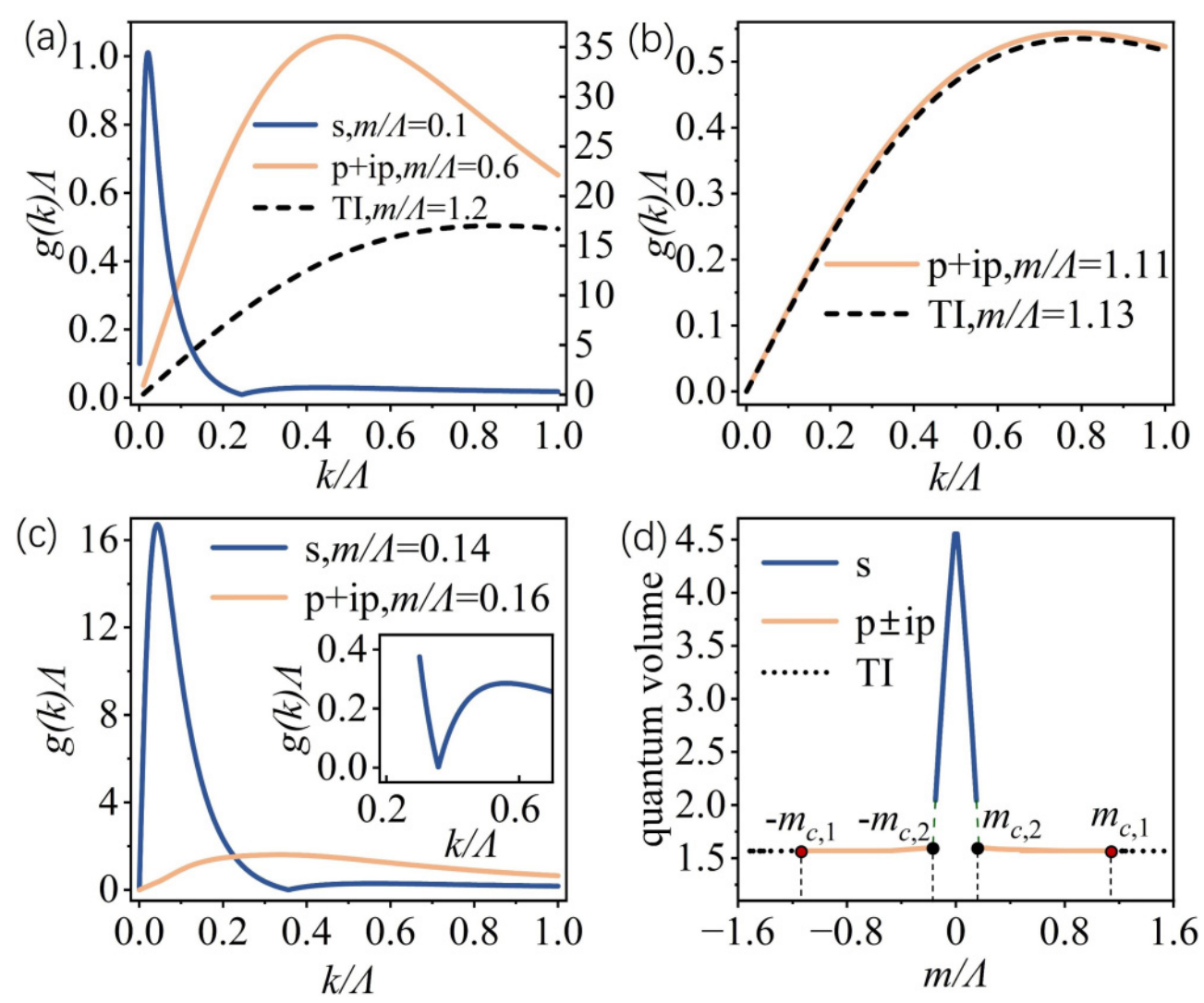

FIG. 3. Comparison of the quantum geometry of three topological insulators. (a) The $k$ distribution of the quantum volume $g(k)$. The left (right) axis corresponds to the $p + ip$ EI and the TI ($s$-wave EI). (b) The change of $g(k)$ crossing the $p + ip$ EI-TI transition (at $m_{c,1} = 1.12$ in Fig. 2). (c) The change of $g(k)$ crossing the $s$-$p$-wave FTEI transition at $m_{c,2} = 0.15$. (d) The evolution of quantum volume with varying $m$.

Appendix E. Notably, the Chern number for both the $s$- and $p + ip$-wave Floquet EI maintain the same as that of the initial Floquet TI, i.e., $C = -1/2$. With further increasing $A$, $m$ changes its sign, three mass-inverted states are found, including the $s$ wave, the $p - ip$-wave EI, and the TI, as shown in Fig. 2. These are the time-reversal symmetry (TRS)-inverted counterparts for those found in $m > 0$, and they exhibit the opposite Chern number $C = 1/2$. The results in Fig. 2(b) clearly show two successive transitions with tuning $m$ (or $A$), i.e., from the Floquet TI to the $p \pm ip$ EI (at $\pm m_{c,1}$) and then to the $s$-wave EI (at $\pm m_{c,2}$). Furthermore, we remark that although a local Hubbardlike interaction is considered here, similar results can be obtained for long-range interactions, as shown by the details in Appendix G.

The single-particle Hamiltonian of the TI can be written compactly as $H_{\mathrm{TI}} = \boldsymbol{\sigma} \cdot \mathbf{d}_{\mathrm{TI}}$, where $\mathbf{d}_{\mathrm{TI}} = (k_x, k_y, m)$. The mean-field Hamiltonians of the $s$- and $p + ip$-wave EI are then accordingly obtained as $H_{\mathrm{EI},s/p} = \boldsymbol{\sigma} \cdot (\mathbf{d}_{\mathrm{TI}} + \delta\mathbf{d}_{s/p})$, where $\delta\mathbf{d}_s = (\frac{m\Delta_s k_x}{\epsilon_k^2}, \frac{m\Delta_s k_y}{\epsilon_k^2}, -\frac{\Delta_s k^2}{\epsilon_k^2})$ and $\delta\mathbf{d}_p = (\frac{\Delta_{p,1}(k_y^2+m^2+m\epsilon_k)}{2\epsilon_k^2}, -\frac{\Delta_{p,1}k_x k_y}{2\epsilon_k^2}, -\frac{\Delta_{p,1}k_x(m+\epsilon_k)}{2\epsilon_k^2})$. Interestingly, although TRS is further broken by both $\delta\mathbf{d}_s$ and $\delta\mathbf{d}_p$, the way how it is broken is different. For the $s$-wave case, TRS is violated by the $z$ component $\delta d_{s,z}$. However, for the $p + ip$ wave case, it is violated by the in-plane components $\delta d_{s,x}$ and $\delta d_{s,y}$. Such a different symmetry property leads to distinct energy spectra shown in Figs. 2(c)–2(e).

Given that both the $s$- and $p + ip$-wave FTEIs exhibit the same Chern number, their physical differences are not characterized by topology. We therefore examine their distinctions regarding the quantum geometry. The geometric information is contained in the Fubini-Study tensor [9,53], i.e., $\mathbb{B}_{ij}(\mathbf{k}) = \langle \partial_i u_{\mathbf{k}} | \partial_j u_{\mathbf{k}} \rangle - \langle \partial_i u_{\mathbf{k}} | u_{\mathbf{k}} \rangle \langle u_{\mathbf{k}} | \partial_j u_{\mathbf{k}} \rangle$, where $|u_{\mathbf{k}}\rangle$ is the wave function defined in the momentum space and $\partial_i = \frac{\partial}{\partial k_i}$ with $i = x, y$. Our main focus is the real part of $\mathbb{B}_{ij}$, i.e., the quantum metric measuring the distance between adjacent quantum states.

To better quantify the quantum geometry for the three phases, we calculate their quantum volumes, defined as $g = \int \mathbf{k}\sqrt{\det \mathrm{Re}[\mathbb{B}_{ij}(\mathbf{k})]}$, as well as the radial distribution of the quantum volume $g(k) = \int_0^{2\pi} d\phi_{\mathbf{k}} k \sqrt{\det \mathrm{Re}[\mathbb{B}_{ij}(\mathbf{k})]}$ [54,55]. Moreover, the relation $\sqrt{\det \mathrm{Re}[\mathbb{B}_{ij}(\mathbf{k})]} = \frac{|\Omega_{xy}|}{2}$ generally holds for two-band models [54]. Consequently, the quantum volume measures the integral of the positive (negative) Berry curvature, depending on whether the total Chern number of the band is negative (positive) [51]. As shown in Fig. 3(a), $g(k)$ generally increases with $k$ for the TI parent state. For the $p + ip$-Floquet EI, $g(k)$ is enhanced and exhibits a wide peak feature. For the $s$-wave EI, however, $g(k)$ undergoes significant changes, displaying a more complicated nonmonotonic behavior. A rapid increase occurs for small $k$ before decreasing to $g(k_0) = 0$, which again increases slightly for larger $k$.

We further examine the change of $g(k)$ crossing the two transition points $m_{c,1}$ and $m_{c,2}$. As shown in Fig. 3(b), $g(k)$ evolves smoothly crossing the $p + ip$ EI-TI transition at $m_{c,1}$. However, an abrupt change occurs at the $p + ip$ EI-$s$-wave EI transition at $m_{c,2}$ [Fig. 3(c)]. Note that $g(k)$ touches zero for the $s$-wave EI. We now briefly discuss the physical meaning of the zeros of the quantum volume, i.e., the point where $g(k_0) = 0$ in Fig. 3(c). This means that $\frac{\partial \psi}{\partial k}|_{k=k_0} = 0$ is satisfied, where $\psi$ is the wave function of the EI. Thus, the radial derivative of the wave function with respect to $k$ vanishes, resulting in similar adjacent quantum states at $k_0$. According to the relation between quantum volume and Berry curvature discussed above, at the point where $k = k_0$, the Berry curvature equals zero and changes of its sign here in momentum space. As discussed above, the sign change leads to physical differences between the quantum volume and the Chern number. Furthermore, the evolution of the quantum volume $g$ with varying $m$ also reveals a sudden jump at $m_{c,2}$ [Fig. 3(d)]. These findings suggest that the $p+ip$ EI shares similar wave function characteristics and quantum geometry with the TI state, while in sharp difference with the $s$-wave EI. As $|m|$ decreases (enhancing the correlation effect), the wave function would intend to maintain the minimal change during phase transitions. This accounts for why the TI first undergoes the transition to the $p+ip$ EI, and then to the $s$-wave EI, consistent with the energetics shown in Figs. 2(a) and 2(b).

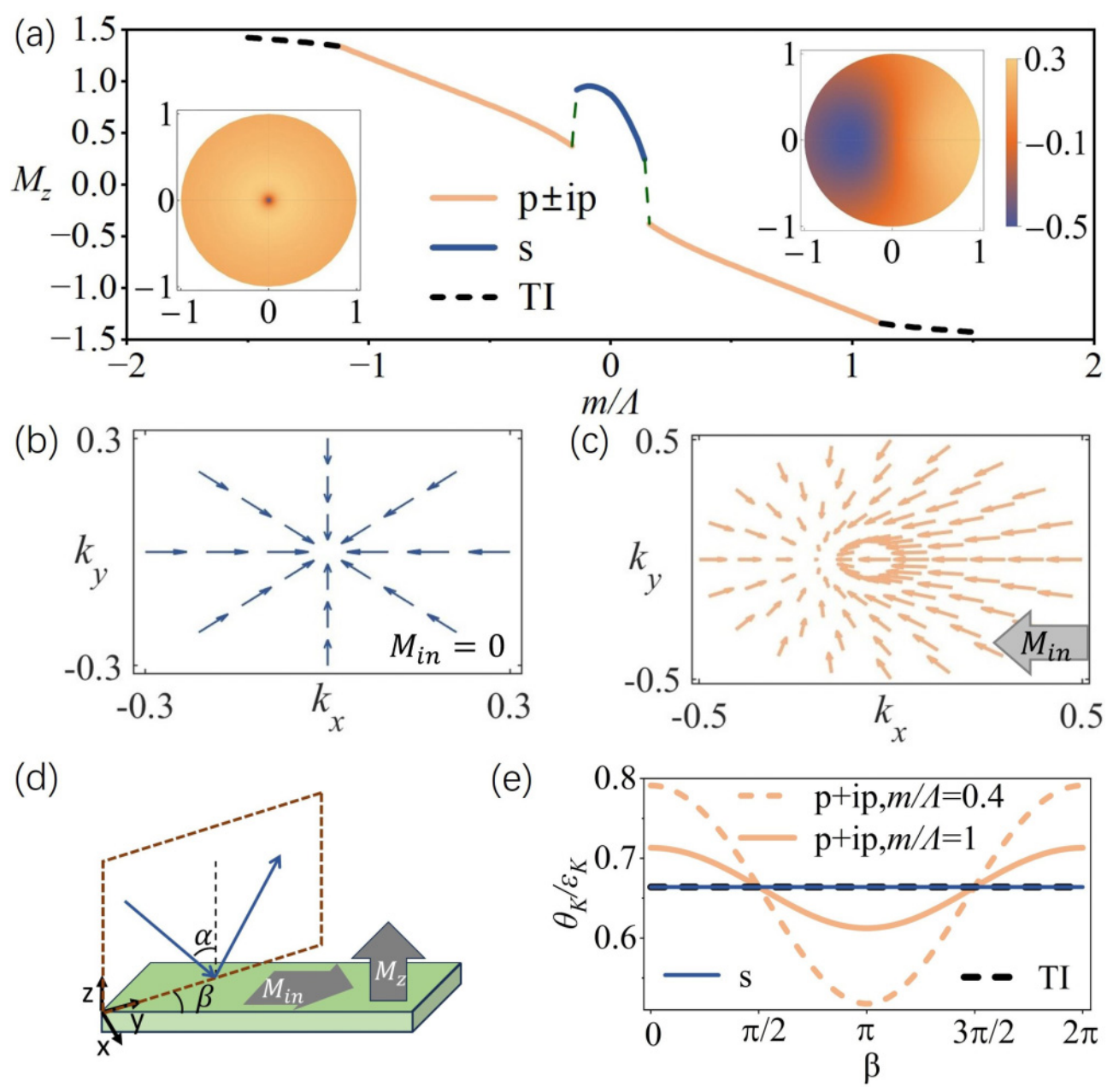

FIG. 4. (a) The magnetization $M_z$ with tuning $m$. The left (right) inset shows the distribution $s_z(\mathbf{k})$ over momentum space for the $s$-wave ($p+ip$-wave) EI. (b), (c) The in-plane spin textures for the $s$-wave and $p+ip$-wave EI. (d) Schematic plot of the setup that demonstrates the MOKE. (e) The ratio between the Kerr angle and ellipticity vs the angle of the incident plane.

## IV. SPIN TEXTURES

So far, we have revealed three Floquet insulators with the same topology but different quantum geometries. The distinct geometries are expected to lead to different physical properties. One direct reflection is the spin texture of ground state, defined as $s_i(\mathbf{k}) = \langle\sigma_i\rangle/2$ $(i = x, y, z)$, where $\langle\cdots\rangle$ denotes the average over the mean-field ground state of the three phases, $|\phi_\mathbf{k}\rangle$. We plot the total spin polarization along $z$, i.e., $M_z = \int d\mathbf{k} s_z(\mathbf{k})$ in Fig. 4(a). As shown, with decreasing $|m|$, the evolution of $S_z$ is smooth during the transition from the TI to the $p \pm ip$ EI. This is due to the fact that $\delta d_{p,z}$ does not violate TRS. In contrast, $M_z$ exhibits a sudden jump at the transition from the $p \pm ip$ EI to the $s$-wave EI, reflecting that TRS is further broken by $\delta d_{s,z}$.

Interestingly, the two EIs also display completely different in-plane spin textures, $\mathbf{s}_{\text{in}} = (s_x(\mathbf{k}), s_y(\mathbf{k}))$, as shown in Figs. 4(b) and 4(c). An anisotropic spin texture is found for the $p+ip$ case, in contrast to the isotropic texture for the $s$-wave EI. Thus, a finite in-plane magnetization, $\mathbf{M}_{\text{in}} = (M_x, M_y) = \int d\mathbf{k}\mathbf{s}_{\text{in}}(\mathbf{k})$, emerges for the $p+ip$ EI, absent in the $s$-wave case. Such characteristic magnetizations could be probed via the measurement of spin susceptibility [56].

## V. MAGNETO-OPTICAL KERR RESPONSES

Magneto-optical Kerr Effect (MOKE) is usually classified into the polar MOKE (P-MOKE), longitudinal MOKE (L-MOKE), and the transversal MOKE (T-MOKE), as illustrated in Fig. 5. For the P-MOKE, the magnetization is perpendicular to the interface and parallel to the plane of incidence. For the L-MOKE, the magnetization is parallel to the interface as well as plane of incidence. For T-MOKE, the magnetization remains parallel to the interface, but perpendicular to the plane of incidence [57]. Since our revealed Floquet topological excitonic insulators exhibit magnetizations along different directions, one expects they display distinct MOKEs in terms of $\theta_K$ and $\epsilon_K$.

The different quantum geometries and magnetizations can also be detected via the MOKE. We consider applying a linear polarized light incident onto the 2D surface states, as indicated in Fig. 4(d). The reflected light will then display both the Kerr rotation $\theta_K$ and the ellipticity $\epsilon_K$, which are determined by the dielectric tensor $\varepsilon$ of the material. For the $s$-wave EI, the magnetization is out of plane ($z$ direction) thus it displays the polar MOKE [58]. The dielectric tensor $\varepsilon$ for this case has $U(1)$ rotation symmetry about the $z$ axis. After a coordinate rotation, the general form of $\varepsilon$ can be obtained for magnetization along arbitrary direction [51,57–59]. Assuming that the electric field vector of the light is perpendicular to the incident plane, $\theta_K$ and $\varepsilon_K$ are derived to satisfy [51]

$$\theta_K + i\varepsilon_K = \frac{in_1n_2Q\cos\alpha_1(\overline{M}_y\sin\alpha_2 + \overline{M}_z\cos\alpha_2)}{\cos\alpha_2(n_2\cos\alpha_1 + n_1\cos\alpha_2)(n_1\cos\alpha_1 - n_2\cos\alpha_2)}, \tag{3}$$

where $\overline{M}_i = M_i/|\mathbf{M}|$ is the normalized magnetization, $n_1$ and $n_2$ are, respectively, the complex refractive indices of the vacuum and the material, $\alpha_1$ and $\alpha_2$ are the incident and reflective angle, and $Q$ is the magneto-optical Voigt constant, which can be experimentally measured. Equating the real and imaginary parts in Eq. (3) and inserting the magnetization of the three insulating states, we can then obtain the ratio $\theta_K/\varepsilon_K$.

Note that only the out-of-plane magnetization is present in the ground state of $s$-wave EIs and noninteracting states, which are exactly characterized by the polar geometry depicted in Fig. 5(a). Thus, it is evident that the Kerr signals $\theta_K$ and $\epsilon_K$ remain unchanged as the plane of incidence (indicated by the dashed box) rotates around the $z$ axis. According to Eq. (3), the Kerr rotation and ellipticity satisfy

$$\theta_K + i\epsilon_K = \frac{in_1n_2Q\cos\alpha_1\overline{M}_z}{(n_2\cos\alpha_1 + n_1\cos\alpha_2)(n_1\cos\alpha_1 - n_2\cos\alpha_2)}. \tag{4}$$

To quantify the MOKEs, we assume that linearly polarized incident light is incident from the vacuum onto the TI surface (e.g., $Bi_2Se_3$) with the incident angle $\alpha_1 = 45^\circ$. The complex refractive index of $Bi_2Se_3$ (e.g., two quintuple layered) for 0.5 μm wavelength light is taken $n_2 = 3.6 + 2.5i$ [60]. Using the Snell-Descartes law [57], $n_1 \sin\alpha_1 = n_2 \sin\alpha_2$, $\mathrm{Re}(n_2 \cos\alpha_2) > 0$, we obtain $\sin\alpha_2 = 0.13 - 0.092i$ and $\cos\alpha_2 = 1.0 + 0.012i$. We then discuss the MOKEs for the three Floquet topological insulators, i.e., the TIs, $s$-wave, and $p + ip$-wave topological EIs.

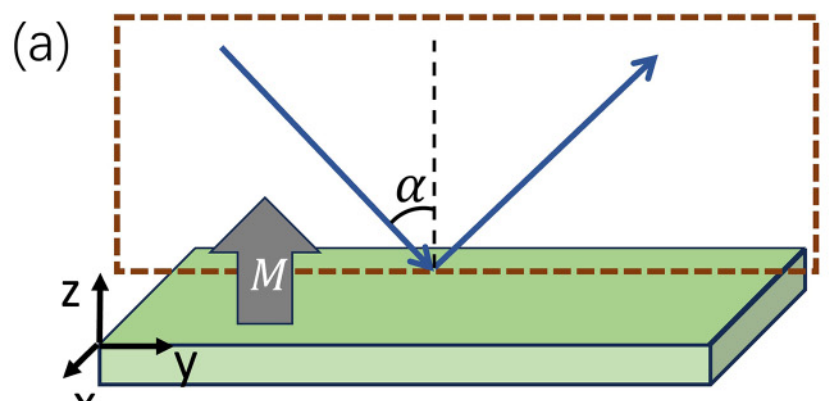

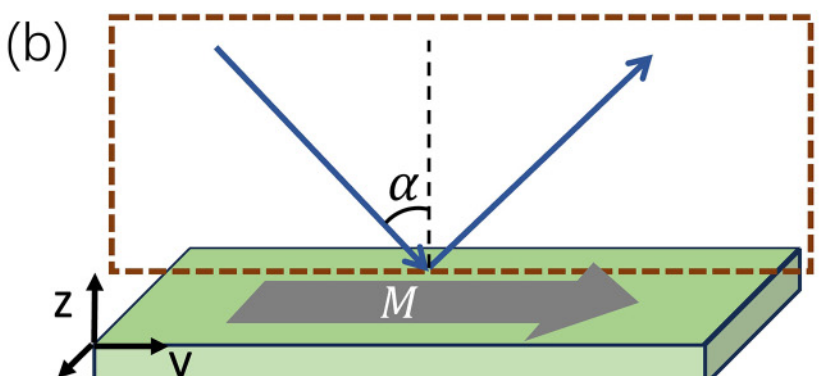

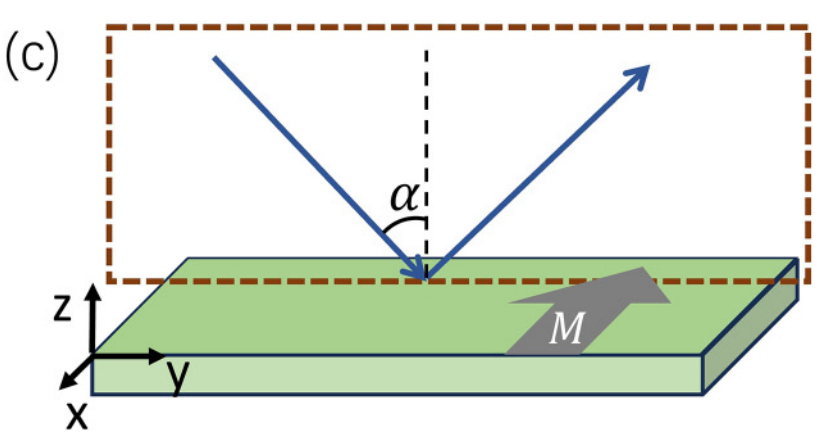

FIG. 5. Schematic plot of three different types of MOKEs. (a) Polar MOKE where magnetization is along $z$ direction. (b) Longitudinal MOKE with the magnetization along $y$ direction (in parallel with the plane of incidence). (c) Transversal MOKE with the magnetization along $y$ direction (perpendicular to the plane of incidence).

For the $s$-wave EI, the magnetization or spin polarization is out-of-plane. As shown in Fig. 4(a), its total magnetization is along the positive $z$ direction, regardless of the sign of $m$. This property is due to the fact that its magnetization is formed via spontaneous symmetry breaking. Notably, this is different from the TI case, in which the out-of-plane magnetization changes between the positive and negative directions along the $z$ axis as the sign of $m$ changes. Thus, for the $s$-wave EI, we always have $\overline{M}_x = \overline{M}_y = 0$ and $\overline{M}_z = M_z/|\mathbf{M}| = 1$. Substituting these into Eq. (4) yields $\theta_K + i\epsilon_K = Q(-0.11 - 0.17i)$. Since the magneto-optical Voigt constant is a parameter yet to be experimentally determined, we treat it as a generic real number and compute the ratio $\theta_K/\epsilon_K$ as the quantity characterizing the MOKE, i.e., $\frac{\theta_K}{\epsilon_K} = 0.66$. For the $s$-wave EI, $\theta_K/\epsilon_K$ is found as a real constant, which neither depends on the strength of magnetization nor on the angle of the incident plane [as a result of the polar symmetry in Fig. 5(a)].

For the TI case, different from the $s$-wave EI, it exhibits positive magnetization along the $z$ direction for $m/\Lambda < 0$ and negative magnetization along the $z$ direction for $m/\Lambda > 0$, as illustrated in Fig. 4(a). For $m/\Lambda < 0$, $\overline{M}_x = \overline{M}_y = 0$, and $\overline{M}_z = M_z/|\mathbf{M}| = 1$, we obtain $\theta_K + i\epsilon_K = Q(-0.11 - 0.17i)$ and $\theta_K/\epsilon_K = 0.66$.

In contrast, for $m/\Lambda > 0, \overline{M}_x = \overline{M}_y = 0$, and $\overline{M}_z = M_z/|\mathbf{M}| = -1$, we arrive at $\theta_K + i\epsilon_K = Q(0.11 + 0.17i)$ and $\theta_K/\epsilon_K = 0.66$. Although the TI and the $s$-wave EI exhibit the same value of $\frac{\theta_K}{\epsilon_K}$, they can in principle be distinguished experimentally. By varying the amplitude of the Floquet fields, $A^2/(\omega\Lambda)$, such that the sign of $m$ is changed, e.g., from $m/\Lambda = 0.1$ to $-0.1$, the Kerr rotation $\theta_K$ and the ellipticity $\epsilon_K$ will be tuned to the opposite values for the $s$-wave EI while remaining intact for the TI.

For the $p + ip$-wave EI, as illustrated in Fig. 4, the ground state exhibits both out-of-plane and in-plane magnetizations, resulting in the coexistence of P-MOKE and L-MOKE/T-MOKE in the system. As we elucidated above, P-MOKE contributes to a constant $\theta_K/\epsilon_K$. However, L-MOKE/T-MOKE would give rise to periodic responses when rotating the plane of incidence, i.e., $\beta$ in Fig. 4(d).

The in-plane magnetization $\mathbf{M}_{\text{in}}$ can be decomposed into components parallel and perpendicular to the plane of incidence, i.e.,

$$\mathbf{M}_{\text{in}} = M_{\text{in}} \cos\beta \mathbf{e}_{\parallel} + +M_{\text{in}} \sin\beta \mathbf{e}_{\perp}. \tag{5}$$

The component $M_{\text{in}} \cos\beta \mathbf{e}_{\parallel}$ leads to L-MOKE, while $M_{\text{in}} \sin\phi \mathbf{e}_{\perp}$ generates the T-MOKE, as indicated by Figs. 5(b) and 5(c). Clearly, as we rotate the plane of incidence ($\beta$), the two components exhibit alternating contribution weights to the MOKE (with the period $2\pi$), driving a periodic response behavior.

Specifically, for the $p + ip$ EI, we have $\overline{M}_y = M_{\text{in}} \cos\phi/\sqrt{|\mathbf{M}_z|^2 + |\mathbf{M}_{\text{in}}|^2}$ and $\overline{M}_z = M_z/\sqrt{|\mathbf{M}_z|^2 + |\mathbf{M}_{\text{in}}|^2}$ in Eq. (3). At $m/\Lambda = -0.4$, we obtain $M_y = M_z = 1.35$. Inserting into Eq. (3), we obtain $\theta_K/\epsilon_K = (-0.022 \cos\beta - 0.081)/(-0.0084 \cos\beta - 0.12)$.

For $m/\Lambda = 1$, the calculated spin texture generates $M_z = -2.46$ and $M_x = -0.91$, leading to $\theta_K/\epsilon_K = (0.011 \cos\beta + 0.11)/(0.0041 \cos\beta + 0.16)$. Clearly, unlike the $s$-wave EI and the TI, the $p + ip$ EI exhibits periodically oscillating MOKE when rotating the plane of incidence. These are direct physical manifestations of their distinct quantum geometries.

We plot $\theta_K/\varepsilon_K$ in Fig. 4(e) by rotating the incident plane, i.e., the angle $\beta$ in Fig. 4(d). As shown, for both the TI and the $s$-wave topological EI, $\theta_K/\varepsilon_K$ remains constant with rotating $\beta$. This is because the TI and the $s$-wave EI only have out-of-plane magnetization $\overline{M}_z$, thus the Kerr response remains rotational invariant with respect to $\beta$. In contrast, the $p + ip$ EI shows a periodic response behavior, whose amplitude decreases with increasing $|m|$. Thus, with tuning the driving field $A$, $\theta_K/\varepsilon_K$ would undergo a nonmonotonic evolution, from a constant behavior to a periodic response, and then back to the constant. The spin texture and the MOKE reveal properties of FTEIs that are not captured by their topology, but are characterized by quantum geometry.

## VI. CONCLUSION AND DISCUSSION

It has been known for decades that $s$-wave EIs can be spontaneously formed for insulators with a gap smaller than the exciton binding energy [21]. However, we show that, if the initial state is a TI, $p \pm ip$ topological EI can emerge as an intermediate state (Fig. 1). Although this goes beyond conventional understanding, it is a natural result from the perspective of quantum geometry. Moreover, the different quantum geometry between the $p \pm ip$ and $s$-wave EIs characterizes essentially different physical properties, such as spin textures and MOKE. In addition, we also show that the insulator gap can be effectively reduced in a nonequilibrium fashion via Floquet engineering. This can significantly enhance the correlation effect, pointing to a promising direction, i.e., exploration of Floquet-correlated phases characterized by quantum geometry.

In addition, we emphasize the difference between our work and several recent studies on similar topics. Reference [61] investigated nonequilibrium excitonic insulators with a time-dependent order parameter. Reference [62] showed that the quantum metric provides an effective tool for probing topological phase transitions. Moreover, Ref. [63] studied exciton induced by optical pumping in the weak-interaction regime, while Ref. [64] explored the emergence of excitonic insulators driven by semimetallic band structures. In contrast, we introduce an approach to optically control the pairing symmetry of excitonic insulators in the strong Coulomb interaction regime, where different phases exhibit distinct quantum geometry and spin textures. To verify these predictions experimentally, the surface of a doped topological insulator might provide a promising platform for observing these physical phenomena. In our calculations, the interaction $U$ is fixed, satisfying $\frac{US\Lambda}{2\pi} = 6\hbar v_F$, where $v_F = 5 \times 10^5$m/s [65] is the Fermi velocity near the Dirac point in topological insulators. Using $\hbar v_F = 0.329$ eV nm and $S \times \Lambda^2 = (2\pi)^2$, we obtain $U = 3.14$ meV. Here, we set the cutoff $\Lambda = 0.1$ Å$^{-1}$ [65]. The value of $\Lambda$ is determined by the range over which the Dirac cone maintains a linear dispersion in the energy spectrum. Thus, our results indicate that when the bare energy gap ($2m$) lies between 0.98 and 7.30 meV, the system exhibits a $p + ip$-wave EI phase, while for gaps smaller than 0.98 meV, the EI phase is of $s$-wave symmetry.

## ACKNOWLEDGMENTS

We acknowledge Tigran Sedrakyan, Qianghua Wang, Junjun Pang, Kai Shao, Yuhan Cao, and Daohe Ma for fruitful discussions. This work was supported by the National Natural Science Foundation of China (No. 12274206 and No. 12322402), the National R&D Program of China (No. 2022YFA1403601 and No. 2024YFA1410500), the Innovation Program for Quantum Science and Technology (No. 2021ZD0302800), the Natural Science Foundation of Jiangsu Province (No. BK20233001), and the Fundamental Research Funds for the Central Universities (Grant No. KG202501).

## DATA AVAILABILITY

The data that support the findings of this article are not publicly available upon publication because it is not technically feasible and/or the cost of preparing, depositing, and hosting the data would be prohibitive within the terms of this research project. The data are available from the authors upon reasonable request.

## APPENDIX A: FLOQUET ENGINEERING OF THE BAND GAP

We resort to the extended Floquet space $\mathcal{F} = \mathcal{H} \otimes \mathcal{T}$, a product of the Hilbert space of the quantum system ($\mathcal{H}$), and the space of square-integrable time-dependent functions with period $T$ ($\mathcal{T}$). The orthonormal complete basis expanding $\mathcal{F}$ is constructed by $|\alpha\nu(t)\rangle = |\alpha\rangle e^{i\nu\omega t}$, where $\nu \in \mathbb{Z}$ and $|\alpha\rangle$ is the many-body basis of $\mathcal{H}$. Meanwhile, the time-dependent Schördinger equation is cast into the eigenvalue problem of the quasienergy operator $Q(t) = H^{\text{tot}}(t) - i\partial_t$. In the extended basis, $Q(t)$ is cast into the matrix $Q_{\alpha'\nu',\alpha\nu} = \langle\alpha'\nu'(t)|Q(t)|\alpha\nu(t)\rangle$ and reads as

$$Q_{\alpha'\nu',\alpha\nu} = \langle\alpha'|H^{\text{tot}}_{\nu'-\nu}|\alpha\rangle + \delta_{\nu'\nu}\delta_{\alpha'\alpha}\nu\omega, \tag{A1}$$

where $H^{\text{tot}}_{\nu} = \frac{1}{T}\int_0^T dt e^{-i\nu\omega t} H^{\text{tot}}(t)$. Evaluating the eigenvalues of $Q_{\alpha'\nu',\alpha\nu}$ is challenging, in particular for our case with many-body interaction $H_{\text{int}}$. We thus formally block diagonalize $Q_{\alpha'\nu',\alpha\nu}$ in the $\mathcal{T}$ space, generating $Q_{\alpha'\alpha,\nu'\nu} = U_{\nu' l} Q_{\alpha'\alpha,l} U^{\star}_{l\nu}$, and $Q_{\alpha'\alpha,l}$ is of the form $Q_{\alpha'\alpha,l} = \langle\alpha'|H_F|\alpha\rangle + \delta_{\alpha'\alpha} l\omega$, where $l$ is the Floquet band index. The Floquet Hamiltonian $H_F$ here is our main focus, as it determines the quasienergy and the stroboscopic evolution of quantum states.

A systematic way to compute $H_F$ is high-frequency expansion [44]. For $\omega$ larger than the off-diagonal components ($\nu' \neq \nu$) in Eq. (A1), the degenerate perturbation calculation can be performed, which generates $H_F = H^{\text{tot}}_0 + [H^{\text{tot}}_1, H^{\text{tot}}_{-1}]/\omega + O(\omega^{-2})$. Here, since the interaction $H_{\text{int}}$ does not have time dependence, its Fourier component is involved in the diagonal component $H^{\text{tot}}_0$. We note that the presence of $H_{\text{int}}$ will introduce further corrections to $H_F$ in the high-frequency expansion. However, the leading corrections do not occur until the third-order perturbation ($\propto O(\omega^{-2})$) [51]. Thus, they can be neglected for high-frequency fields with $\omega \gg u \gtrsim \frac{A^2}{\omega}$ [51].

We start from the Hamiltonian of interacting massive Dirac fermions, i.e.,

$$\begin{aligned} H &= \sum_{\mathbf{k}} \Phi^{\dagger}_{\mathbf{k}}[v_F(k_x\sigma_x + k_y\sigma_y) + u\sigma_z]\Phi_{\mathbf{k}} + H_{\text{int}} \\ &= \sum_{\mathbf{k}} (f^{\dagger}_{\mathbf{k},\uparrow} \quad f^{\dagger}_{\mathbf{k},\downarrow}) \begin{pmatrix} u & v_F(k_x - ik_y) \\ v_F(k_x + ik_y) & -u \end{pmatrix} \begin{pmatrix} f_{\mathbf{k},\uparrow} \\ f_{\mathbf{k},\downarrow} \end{pmatrix} \\ &\quad + U \sum_{\mathbf{k},\mathbf{k}',\mathbf{q}} f^{\dagger}_{\mathbf{k}+\mathbf{q},\uparrow} f^{\dagger}_{\mathbf{k}'-\mathbf{q},\downarrow} f_{\mathbf{k}',\downarrow} f_{\mathbf{k},\uparrow}, \end{aligned} \tag{A2}$$

where $\Phi_{\mathbf{k}} = [f_{\mathbf{k},\uparrow}, f_{\mathbf{k},\downarrow}]^{\text{T}}$ and hereafter we set Fermi velocity $v_F$ to 1. An energy cutoff $\Lambda$ is implicit in the kinetic energy term of Eq. (A2). We then apply the periodic driving field, which drives the kinetic energy, i.e., into the first term in Eq. (A2),

$$H^{\text{kin}}_{\mathbf{k}}(t) = u\sigma_z + (k_x + A\sin\omega t)\sigma_x + (k_y + A\sin(\omega t + \phi))\sigma_y. \tag{A3}$$

We then proceed by performing a Fourier expansion of the total time-dependent Hamiltonian $H^{\text{tot}}(t) = \sum_{\mathbf{k}} \Phi^{\dagger}_{\mathbf{k}} H^{\text{kin}}_{\mathbf{k}}(t)\Phi_{\mathbf{k}} + H_{\text{int}}$, which is cast into

$$H^{\text{tot}} = H^{\text{tot}}_0 + H^{\text{tot}}_1 e^{i\omega t} + H^{\text{tot}}_{-1} e^{-i\omega t}. \tag{A4}$$

Since $H_{\text{int}} = U\sum_{\mathbf{k},\mathbf{k}',\mathbf{q}} f^{\dagger}_{\mathbf{k}+\mathbf{q},\uparrow} f^{\dagger}_{\mathbf{k}'-\mathbf{q},\downarrow} f_{\mathbf{k}',\downarrow} f_{\mathbf{k},\uparrow}$ is time-independent, it is only involved in the component $H^{\text{tot}}_0$. In contrast, the kinetic energy term $H^{\text{kin}}(t) = \sum_{\mathbf{k}} \Phi^{\dagger}_{\mathbf{k}} H^{\text{kin}}_{\mathbf{k}}(t)\Phi_{\mathbf{k}}$ generates all three nonzero components in Eq. (A4). Thus, we obtain

$$\begin{aligned} H^{\text{tot}}_0 &= H^{\text{kin}}_0 + H_{\text{int}} \\ &= \sum_{\mathbf{k}} \Phi^{\dagger}_{\mathbf{k}} \begin{pmatrix} u & k_x - ik_y \\ k_x + ik_y & -u \end{pmatrix} \Phi_{\mathbf{k}} + H_{\text{int}} \end{aligned} \tag{A5}$$

and the Fourier components of Hamiltonian of $H_{\mathbf{k}}^{\text{int}}$ are given by

$$H_1^{\text{tot}} = H_1^{\text{kin}} = \sum_{\mathbf{k}} \Phi_{\mathbf{k}}^{\dagger} \begin{pmatrix} 0 & A\left(\frac{1}{2i} - \frac{e^{i\phi}}{2}\right) \\ A\left(\frac{1}{2i} + \frac{e^{i\phi}}{2}\right) & 0 \end{pmatrix} \Phi_{\mathbf{k}}, \tag{A6}$$

$$\begin{aligned} H_{-1}^{\text{tot}} &= H_{-1}^{\text{kin}} \\ &= \sum_{\mathbf{k}} \Phi_{\mathbf{k}}^{\dagger} \begin{pmatrix} 0 & A\left(-\frac{1}{2i} + \frac{e^{-i\phi}}{2}\right) \\ A\left(-\frac{1}{2i} - \frac{e^{-i\phi}}{2}\right) & 0 \end{pmatrix} \Phi_{\mathbf{k}}. \end{aligned} \tag{A7}$$

In high-frequency expansion, the effective Floquet Hamiltonian can be calculated as

$$\begin{aligned} H_{\text{eff}} = &H_0^{\text{tot}} + \sum_{m\neq 0} \frac{H_m^{\text{tot}} H_{-m}^{\text{tot}}}{m\omega} + \sum_{m\neq 0} \frac{\left[H_{-m}^{\text{tot}}, \left[H_0^{\text{tot}}, H_m^{\text{tot}}\right]\right]}{2(m\omega)^2} \\ &+ O(\omega^{-2}). \end{aligned} \tag{A8}$$

The above expansion holds in the following perturbative sense [44]. The energy scale of $H_0$ is $\sim D = \sqrt{u^2 + \Lambda^2} \sim u$ (because $u = 1.2\Lambda$), and $H_{\text{int}} \sim \Lambda$ is taken in this work. The Floquet field modulates system mainly by the first-order term in Eq. (A8), i.e., $A^2/\omega$. Meanwhile, the second term in Eq. (A8), i.e., $\frac{A^2 u}{\omega^2}$ can be neglected under the condition $u \gtrsim \frac{A^2}{\omega} \gg \frac{A^2 u}{\omega^2}$, which implies $\omega \gg u \gtrsim \frac{A^2}{\omega}$. Clearly, this automatically ensures that the different Floquet bands do not overlap. Notably, it is also known from Eq. (A8) that the four-fermion interaction term does not get renormalized until the second-order correction, which is negligible under the above high-frequency condition.

Up to first-order correction, we arrive at

$$H_{\text{eff}} = \sum_{\mathbf{k}} \Phi_{\mathbf{k},0}^{\dagger} \begin{pmatrix} u + \frac{A^2 \sin\phi}{\omega} & k_x - ik_y \\ k_x + ik_y & -u - \frac{A^2 \sin\phi}{\omega} \end{pmatrix} \Phi_{\mathbf{k},0} + H_{\text{int}}. \tag{A9}$$

The first term of Eq. (A9) gives rise to the energy spectrum $\pm\epsilon_{\mathbf{k}} = \pm\sqrt{k^2 + (u + \frac{A^2}{\omega}\sin(\phi))^2}$.

For circular polarization with $\phi = -\frac{\pi}{2}$, the Floquet Hamiltonian is reduced to that in the main text, with the effective energy gap, $m = u - \frac{A^2}{\omega}$, which is modulated by external driving fields.

## APPENDIX B: THE VALIDITY OF THE PARAMETERS

Our parameter regime is characterized by $\omega \gg u \gtrsim \frac{A^2}{\omega}$, where $u$ is the mass term in noninteracting Hamiltonian and $\Lambda$ is the cutoff of the momentum $k$.

We first explain why it is sufficient to consider only the zeroth Floquet band. The noninteracting Floquet Hamiltonian is given by

$$H_F = \begin{pmatrix} \ddots & H_1 & 0 & 0 & 0 \\ H_{-1} & H_0 + \omega & H_1 & 0 & 0 \\ 0 & H_{-1} & H_0 & H_1 & 0 \\ 0 & 0 & H_{-1} & H_0 - \omega & H_1 \\ 0 & 0 & 0 & H_{-1} & \ddots \end{pmatrix}, \tag{B1}$$

where

$$\begin{aligned} H_0 &= \begin{pmatrix} u & k_x - ik_y \\ k_x + ik_y & -u \end{pmatrix}, \\ H_1 &= \begin{pmatrix} 0 & A\left(\frac{1}{2i} - \frac{e^{i\phi}}{2}\right) \\ A\left(\frac{1}{2i} + \frac{e^{i\phi}}{2}\right) & 0 \end{pmatrix}, \\ H_{-1} &= \begin{pmatrix} 0 & A\left(-\frac{1}{2i} + \frac{e^{-i\phi}}{2}\right) \\ A\left(-\frac{1}{2i} - \frac{e^{-i\phi}}{2}\right) & 0 \end{pmatrix}. \end{aligned} \tag{B2}$$

Here, $H_0 \sim D$, $H_{\pm 1} \sim A$, where $D = \sqrt{u^2 + \Lambda^2}$. The parameter $u$ is chosen to be $1.2\Lambda$, so that $D \sim u$ holds. The interaction $U$ primarily affects low-energy physics and does not significantly modify the bandwidth, so the overall bandwidth remains on the order of $u$. Structurally, this Floquet Hamiltonian resembles a tight-binding model with nearest-neighbor hopping in a linear potential, which is analogous to the problem of an electron in a 1D lattice under a uniform electric field. This analogy gives rise to the Bloch oscillations and the "Wannier-Stark ladder" [66,67].

Given $\omega \gg u \gtrsim \frac{A^2}{\omega}$, we have $\omega \gg u$ and $\omega^2 \gg A^2$, and hence the system exhibits Wannier-Stark localization under this condition. In this regime, electrons in the zeroth Floquet band are strongly suppressed from transitioning to neighboring Floquet bands due to large energy offsets between them. Therefore, only the zeroth Floquet band needs to be retained, with corrections arising from virtual hopping processes. This approximation is widely used [46,68]. In a system driven by multiple high-frequency fields, localization appears near the zero-frequency equipotential line [69].

We now consider the parameter condition $\omega \gg u \gtrsim \frac{A^2}{\omega}$. A general framework for the high frequency expansions of Floquet effective Hamiltonian is provided below [70,71]. Up to second order $O(1/\omega^2)$, the effective Hamiltonian is given by [71]:

$$\begin{aligned} \hat{H}_{\text{eff}} = &\hat{H}_0 + \frac{1}{\omega} \sum_{j=1}^{\infty} \frac{1}{j} [\hat{V}^{(j)}, \hat{V}^{(-j)}] \\ &+ \frac{1}{2\omega^2} \sum_{j=1}^{\infty} \frac{1}{j^2} ([[\hat{V}^{(j)}, \hat{H}_0], \hat{V}^{(-j)}] + \text{H.c.}) \\ &+ \frac{1}{3\omega^2} \sum_{j,l=1}^{\infty} \frac{1}{jl} ([\hat{V}^{(j)}, [\hat{V}^{(l)}, \hat{V}^{(-j-l)}]] \\ &- 2[\hat{V}^{(j)}, [\hat{V}^{(-l)}, \hat{V}^{(l-j)}]] + \text{H.c.}), \end{aligned} \tag{B3}$$

where $\hat{V}^{(j)}$ is the $j$th Fourier component of the Hamiltonian.

In our model, the Floquet effective Hamiltonian reads:

$$\begin{aligned} H_{\text{eff}} = &H_0^{\text{tot}} + \sum_{m\neq 0} \frac{H_m^{\text{tot}} H_{-m}^{\text{tot}}}{m\omega} + \sum_{m\neq 0} \frac{\left[H_{-m}^{\text{tot}}, \left[H_0^{\text{tot}}, H_m^{\text{tot}}\right]\right]}{2(m\omega)^2} \\ &+ O(\omega^{-2}), \end{aligned} \tag{B4}$$

where $H_0^{\text{tot}} = H_0 + H_{\text{int}} \sim u + U$. As we mentioned below, the interaction $U$ does not significantly affect the bandwidth, so $H_0^{\text{tot}} \sim D \sim u$, and $H_{\pm 1}^{\text{tot}} \sim A$. So $= H_0^{\text{tot}} \sim u$, $\sum_{m\neq 0} \frac{H_m^{\text{tot}} H_{-m}^{\text{tot}}}{m\omega} \sim \frac{A^2}{\omega}$, $\sum_{m\neq 0} \frac{[H_{-m}^{\text{tot}}, [H_0^{\text{tot}}, H_m^{\text{tot}}]]}{2(m\omega)^2} \sim \frac{uA^2}{\omega^2}$. This is a perturbative expansion in powers of $1/\omega$, where we retain the first- and second-order terms. To ensure a controlled perturbative expansion, we require $u \gtrsim \frac{A^2}{\omega} \gg \frac{uA^2}{\omega^2}$, or equivalently $\omega \gg u \gtrsim \frac{A^2}{\omega}$. $u \gtrsim \frac{A^2}{\omega}$ ensures the first term is larger than the second term, while $\frac{A^2}{\omega} \gg \frac{uA^2}{\omega^2}$ justifies neglecting the $1/\omega^2$ terms.

To further justify the validity of the perturbative expansion in powers of $1/\omega$, we consider the scaling of higher-order term [70–72]. Since $H_1$ describes the transition of an electron from the zeroth Floquet subspace to the first Floquet subspace, to compute the contribution of this virtual process to the zeroth subspace, $H_{-1}$ is needed to bring the electrons back from the first Floquet subspace to the zeroth Floquet subspace. Therefore, $H_1$ and $H_{-1}$ always appear together in the high-frequency expansion. The third-order terms in the form of $1/\omega^3$ include expressions such as $\frac{[H_{-1}^{\rm tot},[H_0^{\rm tot},[H_0^{\rm tot},H_1^{\rm tot}]]]}{\omega^3} \sim \frac{u^2A^2}{\omega^3}$ and $\frac{[H_{-1}^{\rm tot},[H_1^{\rm tot},[H_{-1}^{\rm tot},H_1^{\rm tot}]]]}{\omega^3} \sim \frac{A^4}{\omega^3}$. In condition of $\omega \gg u \gtrsim \frac{A^2}{\omega}$, we can easily obtain $\frac{uA^2}{\omega^2} \gg \frac{u^2A^2}{\omega^3}$ and $\frac{uA^2}{\omega^2} \gtrsim \frac{A^4}{\omega^3}$, which implies that the $1/\omega^2$ terms are larger than the $1/\omega^3$ terms. In the same way, we can obtain fourth-order terms proportional to $1/\omega^4$ including $\frac{u^3A^2}{\omega^4}$ and $\frac{uA^4}{\omega^4}$, and they are smaller than the $1/\omega^3$ terms under the condition of $\omega \gg u \gtrsim \frac{A^2}{\omega}$. So we here verified the validity of high-frequency expansion and truncation at the order of $1/\omega^2$.

## APPENDIX C: REDUCED HAMILTONIAN OF FLOQUET TOPOLOGICAL EXCITONIC INSULATORS

The kinetic energy term in Eq. (2) is diagonalized via the unitary transformation,

$$\begin{pmatrix} f_{\mathbf{k},0,\uparrow} \\ f_{\mathbf{k},0,\downarrow} \end{pmatrix} = \begin{pmatrix} \cos\theta_{k,m} e^{-i\phi_{\mathbf{k}}} & \sin\theta_{k,m} e^{-i\phi_{\mathbf{k}}} \\ \sin\theta_{k,m} & -\cos\theta_{k,m} \end{pmatrix} \begin{pmatrix} c_{\mathbf{k}} \\ v_{\mathbf{k}} \end{pmatrix}. \tag{C1}$$

where $c_{\mathbf{k}}$ and $v_{\mathbf{k}}$ denote the annihilation operator for the Floquet modes in the conduction and valence band, and $\phi_{\mathbf{k}}$ is the angle of $\mathbf{k}$, and $\sin\theta$ and $\cos\theta$ are functions solely determined by $k$ and $m$. For simplicity, we abbreviate $\sin\theta_{k,m}$ and $\cos\theta_{k,m}$ as $\sin\theta_k$ and $\cos\theta_k$, respectively.

We insert the unitary transformation into the interaction term $H_{\rm int} = U\sum_{\mathbf{k},\mathbf{k}',\mathbf{q}} f^\dagger_{\mathbf{k}+\mathbf{q},0,\uparrow} f^\dagger_{\mathbf{k}'-\mathbf{q},0,\downarrow} f_{\mathbf{k}',0,\downarrow} f_{\mathbf{k},0,\uparrow}$ in Eq. (2), 16 different terms are derived describing the scattering between electrons on the conduction and valence bands. Focusing on the terms with energy/momentum conservation and the interband interactions in the forward scattering channel (that favors excitonic instabilities), we can reduce the interaction term to the form tractable by mean-field treatment. For the scattering channel, $\mathbf{k}+\mathbf{q}=\mathbf{k}$, i.e., $\mathbf{q}=\mathbf{0}$, we obtain

$$\begin{aligned} H_{\rm int}^{(1)} = {}& U\sum_{\mathbf{k},\mathbf{k}'}[(\cos^2\theta_k\cos^2\theta_{k'} + \sin^2\theta_k\sin^2\theta_{k'})c^\dagger_{\mathbf{k}} v^\dagger_{\mathbf{k}'} v_{\mathbf{k}'} c_{\mathbf{k}} \\ & + 2\cos\theta_k\cos\theta_{k'}\sin\theta_k\sin\theta_{k'} c^\dagger_{\mathbf{k}} v^\dagger_{\mathbf{k}'} v_{\mathbf{k}} c_{\mathbf{k}'}]. \end{aligned} \tag{C2}$$

For the other scattering channel $\mathbf{k}+\mathbf{q}=\mathbf{k}'$, we arrive at

$$\begin{aligned} H_{\rm int}^{(2)} = {}& U\sum_{\mathbf{k},\mathbf{k}'}[2\cos(\phi_{\mathbf{k}}-\phi_{\mathbf{k}'})\cos\theta_k\cos\theta_{k'}\sin\theta_k\sin\theta_{k'} \\ & \times c^\dagger_{\mathbf{k}} v^\dagger_{\mathbf{k}'} v_{\mathbf{k}'} c_{\mathbf{k}}] + U\sum_{\mathbf{k},\mathbf{k}'}\big[\cos^2\theta_k\cos^2\theta_{k'} e^{i(\phi_{\mathbf{k}}-\phi_{\mathbf{k}'})} \\ & + \sin^2\theta_k\sin^2\theta_{k'} e^{i(\phi_{\mathbf{k}'}-\phi_{\mathbf{k}})}\big] c^\dagger_{\mathbf{k}} v^\dagger_{\mathbf{k}'} v_{\mathbf{k}} c_{\mathbf{k}'}. \end{aligned} \tag{C3}$$

We note that at the mean-field level, the first term in Eqs. (C2) and (C3) only generates a shift of the chemical potential, which is negligible for insulating phases where the Fermi energy lies in the band gap. Thus, combining $H_{\rm int}^{(1)}$ and $H_{\rm int}^{(2)}$, we obtain the reduced interaction relevant to the Floquet excitonic insulator phases, i.e., Eq. (3).

## APPENDIX D: THE SELF-CONSISTENT SOLUTIONS

(1) When $\mathbf{k}+\mathbf{q}=\mathbf{k}$, i.e., $\mathbf{q}=\mathbf{0}$, mean-field interband Hamiltonian can be written as

$$\begin{aligned} \bar{H}_{\rm int} = {}& U\sum_{\mathbf{k},\mathbf{k}'}[(\cos^2\theta_k\cos^2\theta_{k'} + \sin^2\theta_k\sin^2\theta_{k'})(\langle c^\dagger_{\mathbf{k}} c_{\mathbf{k}}\rangle v^\dagger_{\mathbf{k}'} v_{\mathbf{k}'} + c^\dagger_{\mathbf{k}} c_{\mathbf{k}}\langle v^\dagger_{\mathbf{k}'} v_{\mathbf{k}'}\rangle - \langle c^\dagger_{\mathbf{k}} c_{\mathbf{k}}\rangle\langle v^\dagger_{\mathbf{k}'} v_{\mathbf{k}'}\rangle) \\ & - 2\cos\theta_k\cos\theta_{k'}\sin\theta_k\sin\theta_{k'}(\langle c^\dagger_{\mathbf{k}} v_{\mathbf{k}}\rangle v^\dagger_{\mathbf{k}'} c_{\mathbf{k}'} + c^\dagger_{\mathbf{k}} v_{\mathbf{k}}\langle v^\dagger_{\mathbf{k}'} c_{\mathbf{k}'}\rangle - \langle c^\dagger_{\mathbf{k}} v_{\mathbf{k}}\rangle\langle v^\dagger_{\mathbf{k}'} c_{\mathbf{k}'}\rangle)]. \end{aligned} \tag{D1}$$

and

$$\bar{H} = \sum_{\mathbf{k}} \epsilon_k (c^\dagger_{\mathbf{k}} c_{\mathbf{k}} - v^\dagger_{\mathbf{k}} v_{\mathbf{k}}) + \bar{H}_{\rm int}. \tag{D2}$$

We note that $\langle c^\dagger_{\mathbf{k}} c_{\mathbf{k}}\rangle$ and $\langle v^\dagger_{\mathbf{k}} v_{\mathbf{k}}\rangle$ are the electron occupations in the conduction bands $n_{c,\mathbf{k}}$ and the valence bands $n_{v,\mathbf{k}}$ separately. Defining the spinor $\Psi^\dagger_{\mathbf{k}} = (c^\dagger_{\mathbf{k}} \quad v^\dagger_{\mathbf{k}})$, we can express the mean-field Hamiltonian in matrix form as

$$\begin{aligned} \bar{H} = {}& \sum_{\mathbf{k}} \Psi_{\mathbf{k}} \begin{pmatrix} \epsilon_k + U(\cos^2\theta_k\cos^2\theta_{k'} + \sin^2\theta_k\sin^2\theta_{k'})n_{v,\mathbf{k}} & -2U\cos\theta_k\cos\theta_{k'}\sin\theta_k\sin\theta_{k'}\langle v^\dagger_{\mathbf{k}'} c_{\mathbf{k}'}\rangle \\ -2U\cos\theta_k\cos\theta_{k'}\sin\theta_k\sin\theta_{k'}\langle c^\dagger_{\mathbf{k}'} v_{\mathbf{k}'}\rangle & -\epsilon_k + U(\cos^2\theta_k\cos^2\theta_{k'} + \sin^2\theta_k\sin^2\theta_{k'})n_{c,\mathbf{k}} \end{pmatrix} \Psi_{\mathbf{k}} \\ & - U\sum_{\mathbf{k},\mathbf{k}'}(\cos^2\theta_k\cos^2\theta_{k'} + \sin^2\theta_k\sin^2\theta_{k'})n_{c,\mathbf{k}} n_{v,\mathbf{k}} + 2U\sum_{\mathbf{k},\mathbf{k}'}\cos\theta_k\cos\theta_{k'}\sin\theta_k\sin\theta_{k'}\langle c^\dagger_{\mathbf{k}} v_{\mathbf{k}}\rangle\langle v^\dagger_{\mathbf{k}'} c_{\mathbf{k}'}\rangle. \end{aligned} \tag{D3}$$

The term $-U\sum_{\mathbf{k},\mathbf{k}'}(\cos^2\theta_k\cos^2\theta_{k'} + \sin^2\theta_k\sin^2\theta_{k'})n_{c,\mathbf{k}} n_{v,\mathbf{k}}$ is equal to a constant, so we ignore it. Diagonal elements $U(\cos^2\theta_k\cos^2\theta_{k'} + \sin^2\theta_k\sin^2\theta_{k'})n_{v,\mathbf{k}}$ and $U(\cos^2\theta_k\cos^2\theta_{k'} + \sin^2\theta_k\sin^2\theta_{k'})n_{c,\mathbf{k}}$ renomalize the Feimi velocity of conduction band and valence band, which do not contribute to excitonic condensation. Hence, it is also neglected within the mean-field approximation. The resulting simplified Hamiltonian is given by

$$\bar{H} = \sum_{\mathbf{k}} \Psi_{\mathbf{k}} \begin{pmatrix} \epsilon_k & -2U\cos\theta_k\cos\theta_{k'}\sin\theta_k\sin\theta_{k'}\langle v^{\dagger}_{\mathbf{k}'}c_{\mathbf{k}'}\rangle \\ -2U\cos\theta_k\cos\theta_{k'}\sin\theta_k\sin\theta_{k'}\langle c^{\dagger}_{\mathbf{k}'}v_{\mathbf{k}'}\rangle & -\epsilon_k \end{pmatrix} \Psi_{\mathbf{k}} + 2U\sum_{\mathbf{k},\mathbf{k}'}\cos\theta_k\cos\theta_{k'}\sin\theta_k\sin\theta_{k'}\langle c^{\dagger}_{\mathbf{k}}v_{\mathbf{k}}\rangle\langle v^{\dagger}_{\mathbf{k}'}c_{\mathbf{k}'}\rangle. \tag{D4}$$

Then, we select $\Delta = U\sum_{\mathbf{k}'}\cos\theta_{k'}\sin\theta_{k'}\langle c^{\dagger}_{\mathbf{k}'}v_{\mathbf{k}'}\rangle$ and the mean-field Hamiltonian can be written as

$$\bar{H} = \sum_{\mathbf{k}} \Psi^{\dagger}_{\mathbf{k}} \begin{pmatrix} \epsilon_k & -2\cos\theta_k\sin\theta_k\Delta^* \\ -2\cos\theta_k\sin\theta_k\Delta & -\epsilon_k \end{pmatrix} \Psi_{\mathbf{k}} + 2\frac{|\Delta|^2}{U} = \sum_{\mathbf{k}} \xi_k(a^{\dagger}_{\mathbf{k}}a_{\mathbf{k}} - b^{\dagger}_{\mathbf{k}}b_{\mathbf{k}}) + 2\frac{|\Delta|^2}{U}, \tag{D5}$$

where $\xi_k = \sqrt{E_0^2(k) + |2\cos\theta_k\sin\theta_k\Delta|^2}$ is the energy of the quasiparticles analogous to that of Bogoliubov quasiparticles. It should be noted that $\Delta$ that is independent of the momentum $\mathbf{k}$ is not the order parameter, whereas $\Delta_{\mathbf{k}} = 2\cos(\theta)_k\sin(\theta)_k\Delta$ defines a momentum-dependent order parameter. Upon the substitution of $\alpha_k = a_k$ and $\beta_k = b^{\dagger}_k$, the above equation is transformed into

$$\bar{H} = \sum_{\mathbf{k}} \xi_k(\alpha^{\dagger}_{\mathbf{k}}\alpha_{\mathbf{k}} + \beta^{\dagger}_{\mathbf{k}}\beta_{\mathbf{k}}) - \sum_{\mathbf{k}} \xi_k + 2\frac{|\Delta|^2}{U}. \tag{D6}$$

In this situation, the ground state energy $E_G$ corresponds to the state of none quasiparticle represented by $\alpha^{\dagger}_{\mathbf{k}}\alpha_{\mathbf{k}} = 0$ and $\beta^{\dagger}_{\mathbf{k}}\beta_{\mathbf{k}} = 0$, i.e.,

$$E_G = -\sum_{\mathbf{k}} \xi_k + 2\frac{|\Delta|^2}{U}. \tag{D7}$$

We can use the variational method to determine the minimum value of the ground state energy, which means

$$\begin{aligned} &\frac{\delta E_G}{\delta\Delta^*} = 0 \\ &-\sum_{\mathbf{k}}\frac{\delta\xi_k}{\delta\Delta^*} + 2\frac{\Delta}{U} = 0. \end{aligned} \tag{D8}$$

We obtain self-consistent equation of $\Delta$ as follows:

$$1 = U\sum_{\mathbf{k}}\frac{\cos^2\theta_k\sin^2\theta_k}{\xi_k}. \tag{D9}$$

To solve this equation, we need to convert the sum to an integral. In this 2D system, this conversion is depicted as

$$\sum_{\mathbf{k}} = \frac{S}{(2\pi)^2}\int kdkd\varphi_{\mathbf{k}}, \tag{D10}$$

where $S$ is the area of the material. Then Eq. (D9) can be rewritten as

$$\frac{S\Lambda U}{(2\pi)}\int_0^1 \frac{k}{\Lambda}d\left(\frac{k}{\Lambda}\right)\frac{\cos^2\theta_k\sin^2\theta_k}{\sqrt{\left(\frac{k}{\Lambda}\right)^2 + \left(\frac{m}{\Lambda}\right)^2 + 4\cos^2\theta_k\sin^2\theta_k\left(\frac{\Delta}{\Lambda}\right)^2}} = 1, \tag{D11}$$

where $\Lambda$ is the cutoff in momentum $k$ introduced to restrict the analysis to the low-energy window. Taking notice of

$$\begin{aligned} \cos\theta_{k,m} &= \frac{k}{\sqrt{k^2 + (\epsilon_k - m)^2}} = \frac{\frac{k}{\Lambda}}{\sqrt{\left(\frac{k}{\Lambda}\right)^2 + \left(\sqrt{\left(\frac{k}{\Lambda}\right)^2 + \left(\frac{m}{\Lambda}\right)^2} - \frac{m}{\Lambda}\right)^2}} = \cos\theta_{\frac{k}{\Lambda},\frac{m}{\Lambda}} \\ \sin\theta_{k,m} &= \frac{\epsilon_k - m}{\sqrt{k^2 + (\epsilon_k - m)^2}} = \frac{\sqrt{\left(\frac{k}{\Lambda}\right)^2 + \left(\frac{m}{\Lambda}\right)^2} - \frac{m}{\Lambda}}{\sqrt{\left(\frac{k}{\Lambda}\right)^2 + \left(\sqrt{\left(\frac{k}{\Lambda}\right)^2 + \left(\frac{m}{\Lambda}\right)^2} - \frac{m}{\Lambda}\right)^2}} = \sin\theta_{\frac{k}{\Lambda},\frac{m}{\Lambda}}. \end{aligned} \tag{D12}$$

It is straightforward to see that $\cos\theta_k$ and $\sin\theta_k$ are functions of $\frac{k}{\Lambda}$ and $\frac{m}{\Lambda}$. The above equations have only two variables, $\frac{\Delta}{\Lambda}$ and $\frac{m}{\Lambda}$, which give rise to a functional relation : : $\frac{\Delta}{\Lambda} = f(\frac{m}{\Lambda})$. In addition, we select $\frac{S\Lambda U}{(2\pi)} = 6$ as a fixed constant. This choice is valid because

$$\Lambda^2 \times S \sim (2\pi)^2, \quad \frac{S\Lambda U}{(2\pi)} = \Lambda U \times \frac{2\pi}{\Lambda^2}, \quad U \sim \frac{6}{2\pi}\Lambda. \tag{D13}$$

The ground-state energy can also be written as

$$\frac{E_G U}{\Lambda^2} = -\frac{SU\Lambda}{2\pi}\int_0^1 \left(\frac{k}{\Lambda}\right) d\left(\frac{k}{\Lambda}\right)\sqrt{\left(\frac{k}{\Lambda}\right)^2+\left(\frac{m}{\Lambda}\right)^2+4\cos^2\theta_k\sin^2\theta_k\left(\frac{\Lambda}{\Lambda}\right)^2}+2\left|\frac{\Delta}{\Lambda}\right|^2. \tag{D14}$$

Thus, the condensation energy is represented by

$$\frac{E_c U}{\Lambda^2} = -\frac{SU\Lambda}{2\pi}\int_0 \left(\frac{k}{\Lambda}\right) d\left(\frac{k}{\Lambda}\right)\left(\sqrt{\left(\frac{k}{\Lambda}\right)^2+\left(\frac{m}{\Lambda}\right)^2+4\cos^2\theta_k\sin^2\theta_k\left(\frac{\Lambda}{\Lambda}\right)^2}-\sqrt{\left(\frac{k}{\Lambda}\right)^2+\left(\frac{m}{\Lambda}\right)^2}\right)+2\left|\frac{\Delta}{\Lambda}\right|^2. \tag{D15}$$

The condensation energy is plotted in Fig. 2.

(2) When $\mathbf{k}+\mathbf{q}=\mathbf{k}'$, the mean-field interband Hamiltonian can be written as

$$\begin{aligned}\bar{H}_{\text{int}} = &-U\sum_{\mathbf{k},\mathbf{k}'}(\cos^2\theta_k\cos^2\theta_{k'}e^{i(\varphi_{\mathbf{k}}-\varphi_{\mathbf{k}'})})(\langle c^\dagger_{\mathbf{k}}v_{\mathbf{k}}\rangle v^\dagger_{\mathbf{k}'}c_{\mathbf{k}'}+c^\dagger_{\mathbf{k}}v_{\mathbf{k}}\langle v^\dagger_{\mathbf{k}'}c_{\mathbf{k}'}\rangle-\langle c^\dagger_{\mathbf{k}}v_{\mathbf{k}}\rangle\langle v^\dagger_{\mathbf{k}'}c_{\mathbf{k}'}\rangle)\\ &-U\sum_{\mathbf{k},\mathbf{k}'}(\sin^2\theta_k\sin^2\theta_{k'}e^{i(\varphi_{\mathbf{k}'}-\varphi_{\mathbf{k}})})(\langle c^\dagger_{\mathbf{k}}v_{\mathbf{k}}\rangle v^\dagger_{\mathbf{k}'}c_{\mathbf{k}'}+c^\dagger_{\mathbf{k}}v_{\mathbf{k}}\langle v^\dagger_{\mathbf{k}'}c_{\mathbf{k}'}\rangle-\langle c^\dagger_{\mathbf{k}}v_{\mathbf{k}}\rangle\langle v^\dagger_{\mathbf{k}'}c_{\mathbf{k}'}\rangle).\end{aligned} \tag{D16}$$

We select the following parameters:

$$\Delta_1 = U\sum_{\mathbf{k}'}\cos^2\theta_{k'}e^{i\varphi_{\mathbf{k}'}}\langle c^\dagger_{\mathbf{k}'}v_{\mathbf{k}'}\rangle, \quad \Delta_2 = U\sum_{\mathbf{k}'}\sin^2\theta_{k'}e^{-i\varphi_{\mathbf{k}'}}\langle c^\dagger_{\mathbf{k}'}v_{\mathbf{k}'}\rangle. \tag{D17}$$

Then, the mean-field Hamiltonian can be written in matrix form as follows:

$$\bar{H} = \sum_{\mathbf{k}}\Psi^\dagger_{\mathbf{k}}\begin{pmatrix}\epsilon_k & -\cos^2\theta_k e^{-i\varphi_{\mathbf{k}}}\Delta_1-\sin^2\theta_k e^{i\varphi_{\mathbf{k}}}\Delta_2\\ -\cos^2\theta_k e^{i\varphi_{\mathbf{k}}}\Delta_1^*-\sin^2\theta_k e^{-i\varphi_{\mathbf{k}}}\Delta_2^* & -\epsilon_k\end{pmatrix}\Psi_{\mathbf{k}}+\frac{|\Delta_1|^2}{U}+\frac{|\Delta_2|^2}{U}. \tag{D18}$$

The same method under the condition of $\mathbf{k}+\mathbf{q}=\mathbf{k}$ is applicable to this mean-field Hamiltonian, resulting in

$$\bar{H} = \sum_{\mathbf{k}}\xi_k(\alpha^\dagger_{\mathbf{k}}\alpha_{\mathbf{k}}+\beta^\dagger_{\mathbf{k}}\beta_{\mathbf{k}})-\sum_{\mathbf{k}}\xi_k+\frac{|\Delta_1|^2}{U}+\frac{|\Delta_2|^2}{U}, \tag{D19}$$

where $\xi_k=\sqrt{E_0^2(k)+|\cos^2\theta_k e^{i\varphi_{\mathbf{k}}}\Delta_1+\sin^2\theta_k e^{-i\varphi_{\mathbf{k}}}\Delta_2|^2}$ holds in the above equation. The ground state is defined as the vacuum state of the $\alpha_{\mathbf{k}}$ and $\beta_{\mathbf{k}}$ quasiparticles, satisfying $\alpha^\dagger_{\mathbf{k}}\alpha_{\mathbf{k}}=0$ and $\beta^\dagger_{\mathbf{k}}\beta_{\mathbf{k}}=0$. Therefore, the ground state energy is

$$E_G = -\sum_{\mathbf{k}}\xi_k+\frac{|\Delta_1|^2}{U}+\frac{|\Delta_2|^2}{U}. \tag{D20}$$

Then, we employ the variational method to seek the ground state energy's minimum value,

$$\frac{\delta E_G}{\Delta_1^*}=0, \quad \frac{\delta E_G}{\Delta_2^*}=0, \tag{D21}$$

i.e.,

$$\begin{aligned}&-\sum_{\mathbf{k}}\cos^2\theta_k e^{-i\varphi_{\mathbf{k}}}(\cos^2\theta_k e^{i\varphi_{\mathbf{k}}}\Delta_1+\sin^2\theta_k e^{-i\varphi_{\mathbf{k}}}\Delta_2)/(2\xi_k)+\frac{\Delta_1}{U}=0,\\ &-\sum_{\mathbf{k}}\sin^2\theta_k e^{i\varphi_{\mathbf{k}}}(\cos^2\theta_k e^{i\varphi_{\mathbf{k}}}\Delta_1+\sin^2\theta_k e^{-i\varphi_{\mathbf{k}}}\Delta_2)/(2\xi_k)+\frac{\Delta_2}{U}=0.\end{aligned} \tag{D22}$$

The above equations can be simplified as

$$\Delta_1 = U\sum_{\mathbf{k}}\frac{\cos^4\theta_k\Delta_1+\cos^2\theta_k\sin^2\theta_k e^{-2i\varphi_{\mathbf{k}}}\Delta_2}{2\xi_k}, \quad \Delta_2 = U\sum_{\mathbf{k}}\frac{\sin^2\theta_k\cos^2\theta_k e^{2i\varphi_{\mathbf{k}}}\Delta_1+\sin^4\theta_k\Delta_2}{2\xi_k}, \tag{D23}$$

which is the set of self-consistent equations.

Note that $\Delta_{\mathbf{k}}=\cos^2(\theta)_k e^{-i\varphi_{\mathbf{k}}}\Delta_1+\sin^2(\theta)_k e^{i\varphi_{\mathbf{k}}}\Delta_2$ is the excitonic order parameter. By selecting a momentum cutoff and transform the summation into an integration, we arrive at

$$\begin{aligned}\frac{\Delta_1}{\Lambda} &= \frac{US\Lambda}{(2\pi)^2}\int_0^1\left(\frac{k}{\Lambda}\right)d\left(\frac{k}{\Lambda}\right)\int_0^{2\pi}d\varphi_{\mathbf{k}}\frac{\cos^4\theta_{\frac{k}{\Lambda}}\frac{\Delta_1}{\Lambda}+\cos^2\theta_{\frac{k}{\Lambda}}\sin^2\theta_{\frac{k}{\Lambda}}e^{-2i\varphi_{\mathbf{k}}}\frac{\Delta_2}{\Lambda}}{2\sqrt{\left(\frac{k}{\Lambda}\right)^2+\left(\frac{m}{\Lambda}\right)^2+\left|\cos^2\theta_{\frac{k}{\Lambda}}e^{i\varphi_{\mathbf{k}}}\frac{\Delta_1}{\Lambda}+\sin^2\theta_{\frac{k}{\Lambda}}e^{-i\varphi_{\mathbf{k}}}\frac{\Delta_2}{\Lambda}\right|^2}},\\ \frac{\Delta_2}{\Lambda} &= \frac{US\Lambda}{(2\pi)^2}\int_0^1\left(\frac{k}{\Lambda}\right)d\left(\frac{k}{\Lambda}\right)\int_0^{2\pi}d\varphi_{\mathbf{k}}\frac{\cos^2\theta_{\frac{k}{\Lambda}}\sin^2\theta_{\frac{k}{\Lambda}}e^{2i\varphi_{\mathbf{k}}}\frac{\Delta_1}{\Lambda}+\sin^4\theta_{\frac{k}{\Lambda}}\frac{\Delta_2}{\Lambda}}{2\sqrt{\left(\frac{k}{\Lambda}\right)^2+\left(\frac{m}{\Lambda}big\right)^2+\left|\cos^2\theta_{\frac{k}{\Lambda}}e^{i\varphi_{\mathbf{k}}}\frac{\Delta_1}{\Lambda}+\sin^2\theta_{\frac{k}{\Lambda}}e^{-i\varphi_{\mathbf{k}}}\frac{\Delta_2}{\Lambda}\right|^2}}.\end{aligned} \tag{D24}$$

Under the condition of $\frac{US\Lambda}{2\pi} = 6$, both $\frac{\Delta_1}{\Lambda}$ and $\frac{\Delta_2}{\Lambda}$ are functions of $\frac{m}{\Lambda}$. Specifically, $\Delta_2 = 0$ when $m > 0$ and $\Delta_1 = 0$ when $m < 0$. The two functions are symmetric with respect to $m = 0$.

When $m > 0$,

$$E_G = -\sum_{\mathbf{k}} \xi_k + \frac{|\Delta_1|^2}{U} = -\sum_{\mathbf{k}} \sqrt{k^2 + m^2 + |\cos^2\theta_k e^{i\varphi_{\mathbf{k}}} \Delta_1|^2} + \frac{|\Delta_1|^2}{U}, \tag{D25}$$

we introduce a momentum cutoff $\Lambda$ and convert summation over $\mathbf{k}$ into an integral

$$\frac{E_G U}{\Lambda^2} = -\frac{US\Lambda}{(2\pi)^2} \int_0^1 \left(\frac{k}{\Lambda}\right) d\left(\frac{k}{\Lambda}\right) \int_0^{2\pi} d\varphi_{\mathbf{k}} \sqrt{\left(\frac{k}{\Lambda}\right)^2 + \left(\frac{m}{\Lambda}\right)^2 + \left|\cos^2\theta_{\frac{k}{\Lambda}} e^{i\varphi_{\mathbf{k}}} \frac{\Delta_1}{\Lambda}\right|^2} + \left|\frac{\Delta_1}{\Lambda}\right|^2. \tag{D26}$$

The condensation energy $E_c$ is defined as the difference between the ground state energy and the noninteracting energy, and can be expressed as

$$\frac{E_c U}{\Lambda^2} = -\frac{US\Lambda}{2\pi} \int_0^1 \left(\frac{k}{\Lambda}\right) d\left(\frac{k}{\Lambda}\right) \left(\sqrt{\left(\frac{k}{\Lambda}\right)^2 + \left(\frac{m}{\Lambda}\right)^2 + \cos^4\theta_{\frac{k}{\Lambda}} \left|\frac{\Delta_1}{\Lambda}\right|^2} - \sqrt{\left(\frac{k}{\Lambda}\right)^2 + \left(\frac{m}{\Lambda}\right)^2}\right) + \left|\frac{\Delta_1}{\Lambda}\right|^2. \tag{D27}$$

When $m < 0$, the result is similar due to $\Delta_1 = 0$ and $\Delta_2 \neq 0$.

We note that our results are based on the self-consistent method in $s$-wave EIs and $p + ip$-wave EIs separately. When the self-consistent equations are solved simultaneously, the results are the same.

## APPENDIX E: EFFECTIVE INTERACTION STRENGTHS OF $s$- AND $p + ip$-WAVE EIs

The order parameter

$$\Delta_{\mathbf{k},m} = \sum_{\mathbf{k}'} V_{\mathbf{k},\mathbf{k}',m} \langle c_{\mathbf{k}}^{\dagger} v_{\mathbf{k}} \rangle \tag{E1}$$

is dependent on momentum $\mathbf{k}$ and mass term $m$, which is determined by the extended field. The $s$- and $p + ip$-wave excitonic insulators correspond to different forms of interaction, given by

$$\begin{aligned} V^{s}_{\mathbf{k},\mathbf{k}',m} &= U \sin\theta_k \cos\theta_k \sin\theta_{k'} \cos\theta_{k'}, \\ V^{p+ip}_{\mathbf{k},\mathbf{k}',m} &= U \cos^2\theta_k \cos^2\theta_{k'} e^{i\varphi_{\mathbf{k}'}} e^{-i\varphi_{\mathbf{k}}}. \end{aligned} \tag{E2}$$

Fixing specific values of $\mathbf{k}$ and $\mathbf{k}'$, we analyze how $V^s$ and $V^{p+ip}$ evolve as functions of the mass term $m$. As $m$ gradually increases from 0, $\sin\theta_k \cos\theta_k$ decreases, while $\cos^2\theta_k$ increases. This indicates that $V^s$ decreases as $m$ increases, whereas $V^{p+ip}$ increases.

Exciton condensation is favored by a smaller energy gap $m$ and a stronger effective interaction. In other words, the formation of exciton condensation is governed by the competition between the energy gap and the effective interaction strength. For $s$-wave excitonic insulators (EIs), as $m$ increases, $V^s$ decreases, resulting in fewer condensed $s$-wave excitons and a reduced condensation energy. In contrast, for $p + ip$-wave EIs, $V^{p+ip}$ increases with $m$ at first, leading to an initial increase in condensation energy, followed by a decrease due to the growing energy gap. Therefore, the condensation energy of the $p + ip$-wave EIs exhibit a nonmonotonic behavior as a function of $m$.

## APPENDIX F: THE QUANTUM VOLUME

The matrix element of the quantum geometry tensor is given by

$$\mathbb{B}_{ij}(\mathbf{k}) = \langle \partial_i u_{\mathbf{k}} | \partial_j u_{\mathbf{k}} \rangle - \langle \partial_i u_{\mathbf{k}} | u_{\mathbf{k}} \rangle \langle u_{\mathbf{k}} | \partial_j u_{\mathbf{k}} \rangle, \tag{F1}$$

where $\partial_i = \frac{\partial}{\partial_{k_i}}$ with $i = x, y$, and $|u_{\mathbf{k}}\rangle$ denotes the $\mathbf{k}$-dependent wave function. The quantum volume $g$ is defined as

$$g = \int d\mathbf{k} \sqrt{\det \mathrm{Re}[\mathbb{B}_{ij}(\mathbf{k})]}, \tag{F2}$$

where det denotes determinant of the matrix with $i$, $j$ indices. The quantum volume and its radial distribution for three different phases are shown in Fig. 3. Here, we show more details regarding its distribution in momentum space. We plot in Fig. 6 the function $\sqrt{\det \mathrm{Re}[\mathbb{B}_{ij}(\mathbf{k})]}$. As shown, a singularity at $\mathbf{k} = 0$ occurs for the $s$-wave EI cases, in sharp contrast to the smooth landscape for the $p + ip$ EI and TI cases. This reflects the fact that the formation of the $s$-wave EI from a spin-momentum locked TI requires a sharp change in terms of wave functions and underlying quantum geometry.

Thus, the $p + ip$-wave EI naturally occurs as an intermediate state for moderate interactions.

There is a property that holds only for two-band models: $\sqrt{\det(\chi)} = \frac{|\Omega_{xy}|}{2}$ [54]. Here, $\det(\chi)$ is the determinant of the quantum metric tensor and $\Omega_{xy}$ is the Berry curvature of the energy band. The quantum metric $\chi_{ij}$ and the Berry curvature $\Omega_{ij}$ are defined through the real and imaginary parts of the quantum geometric tensor as

$$\begin{aligned} \chi_{ij}(\boldsymbol{\lambda}) &\equiv \mathrm{Re}[\mathbb{B}_{ij}(\boldsymbol{\lambda})], \\ \Omega_{ij}(\boldsymbol{\lambda}) &\equiv -2\,\mathrm{Im}[\mathbb{B}_{ij}(\boldsymbol{\lambda})]. \end{aligned} \tag{F3}$$

If the Berry curvature does not change its sign in the momentum space of our continuum model, the quantum volume coincides with $\pi|C|$, where $C$ is the Chern number. This result coincides with the situation of the noninteracting normal insulators and $p + ip$-wave EIs. As shown in Fig. 3(d), the quantum volumes of normal insulators and $p + ip$-wave EIs are around $1.57 \approx \pi \cdot \frac{1}{2} = \pi \cdot |C|$. In contrast, the Berry curvature changes its sign in momentum space for $s$-wave EIs, resulting in quantum volume $\int \sqrt{\det(\chi)}d\boldsymbol{k} = \int \frac{|\Omega_{xy}|}{2} d\boldsymbol{k} \neq \int \frac{\Omega_{xy}}{2} d\boldsymbol{k} = \pi \cdot C$. Thus, the quantum volume is not proportional to the Chern number $C$ for $s$-wave EIs. For $m > 0$, $g - 1.57 = g - \pi \times |C| = \int d^2\mathbf{k}\frac{|\Omega_{xy}|}{2} + \int d^2\mathbf{k}\frac{\Omega_{xy}}{2} = 2\int d^2\mathbf{k}\frac{\Omega_{xy}^{+}}{2}$. The quantum volume describes the integration of positive Berry curvature when the total Chern number of this band is negative. In contrast, the quantum volume describes the integration of the negative Berry curvature when the total Chern number of this band is positive because $g - 1.57 = -2\int d^2\mathbf{k}\frac{\Omega_{xy}^{-}}{2}$ for $m < 0$. In experiment, there are many ways to measure the Berry curvature) $1.57 \approx \pi \times \frac{1}{2} = \pi \times |C|$ [12,73]. The Berry curvature in noninteracting TI and $p + ip$-wave EI does not change sign, so their quantum volumes equal $\pi|C| \approx 1.57$, which is shown in Fig. 3(b). However, the Berry curvature in $s$-wave EI changes its sign in momentum space, resulting in its quantum volume exceeding 1.57.

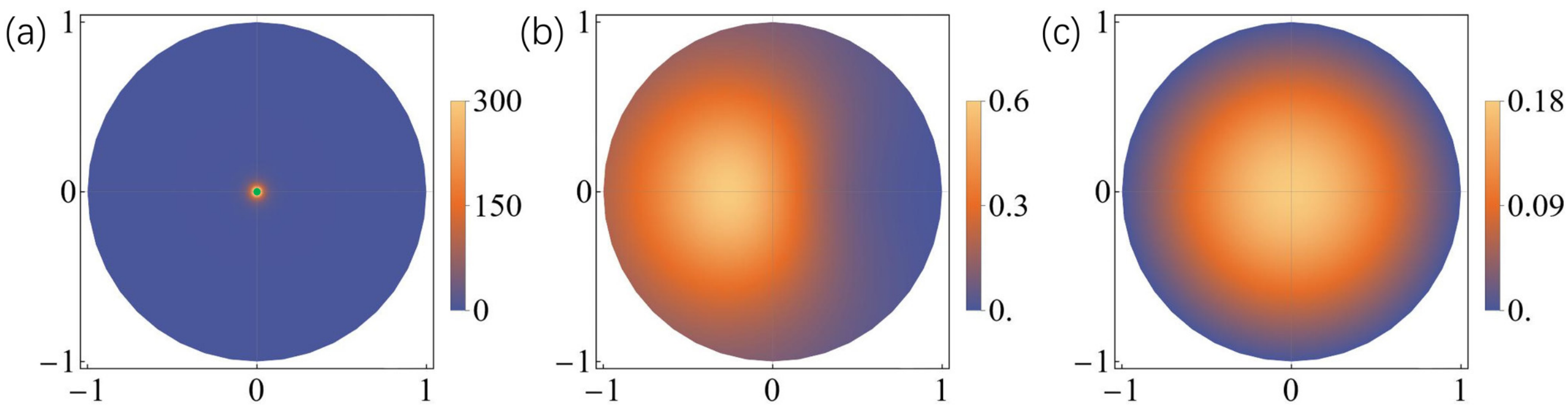

FIG. 6. The distribution of quantum volume in $\mathbf{k}$ space of (a) $s$-EI at $m/\Lambda = 0.1$, (b) $p + ip$-EI at $m/\Lambda = 0.6$, and (c) TI at $m/\Lambda = 1.2$. The green dot in panel (a) indicates a diverging point.

## APPENDIX G: LONG-RANGE INTERACTIONS

The typical long-range interactions characterizing these systems do not affect the results discussed here.

According to Lindhard theory, long-range Coulomb interaction takes the form of [74,75]

$$H_{\text{int}} = \sum_{\mathbf{k},\mathbf{k}',\mathbf{q}} \frac{2\pi e^2}{\epsilon(q+\kappa)} d^{\dagger}_{\mathbf{k}+\mathbf{q}} f^{\dagger}_{\mathbf{k}'-\mathbf{q}} f_{\mathbf{k}'} d_{\mathbf{k}}, \tag{G1}$$

where $\epsilon$ is the permittivity and $\kappa$ is the 2D screening wave number (i.e., 2D inverse screening length), defined as $\kappa = \frac{2\pi e^2}{\epsilon}\frac{\partial n}{\partial \mu}$.

(i) For $\mathbf{k} + \mathbf{q} = \mathbf{k}$, i.e., $\mathbf{q} = 0$, the interband Coulomb interaction can be written as

$$H_{\text{int}} = \frac{2\pi e^2}{\epsilon\kappa} \sum_{\mathbf{k},\mathbf{k}'} (2\cos\theta_k \cos\theta_{k'} \sin\theta_k \sin\theta_{k'} c^{\dagger}_{\mathbf{k}} v^{\dagger}_{\mathbf{k}'} v_{\mathbf{k}} c_{\mathbf{k}}). \tag{G2}$$

Following the same approach used for the Hubbard interaction, we set $\Delta = \frac{2\pi e^2}{\epsilon\kappa}\sum_{\mathbf{k}} \cos\theta_k \sin\theta_k \langle c^{\dagger}_{\mathbf{k}} v_{\mathbf{k}} \rangle$. From this, the following self-consistent equation is obtained

$$1 = \frac{2\pi e^2}{\epsilon\kappa} \sum_{\mathbf{k}} \frac{\cos^2\theta_k \sin^2\theta_k}{\xi_k}, \tag{G3}$$

where $\xi_k = \sqrt{k^2 + m^2 + 4\cos^2\theta_k \sin^2\theta_k \Delta^2}$. We choose a cutoff $\Lambda$ in momentum $k$ and then the self-consistent equation can be written as

$$1 = \frac{S\Lambda}{2\pi} \cdot \frac{2\pi e^2}{\epsilon\kappa} \int_0^1 \frac{k}{\Lambda} d\left(\frac{k}{\Lambda}\right) \frac{\cos^2\theta_{k/\Lambda} \sin^2\theta_{k/\Lambda}}{\sqrt{(k/\Lambda)^2 + (m/\Lambda)^2 + 4\cos^2\theta_{k/\Lambda} \sin^2\theta_{k/\Lambda} (\Delta/\Lambda)^2}}. \tag{G4}$$

We can also calculate the condensation energy by

$$\frac{E_c}{\Lambda^2} \cdot \frac{2\pi e^2}{\epsilon\kappa} = -\frac{S\Lambda}{2\pi} \cdot \frac{2\pi e^2}{\epsilon\kappa} \int_0^1 \frac{k}{\Lambda} d\left(\frac{k}{\Lambda}\right) \left(\sqrt{(k/\Lambda)^2 + (m/\Lambda)^2 + 4\cos^2\theta_k \sin^2\theta_k (\Delta/\Lambda)^2} - \sqrt{(k/\Lambda)^2 + (m/\Lambda)^2}\right) + 2\left|\frac{\Delta}{\Lambda}\right|^2. \tag{G5}$$

Here, we take the parameter $\frac{S\Lambda}{2\pi} \cdot \frac{2\pi e^2}{\epsilon\kappa} = U_0 = 6$.

(2) For $\mathbf{k}+\mathbf{q}=\mathbf{k}'$, the interband mean-field Coulomb interaction is written as

$$\bar{H}_{\text{int}} = -\sum_{\mathbf{k},\mathbf{k}'}\frac{2\pi e^2}{\epsilon(\kappa+|\mathbf{k}-\mathbf{k}'|)}(\cos^2\theta_k\cos^2\theta_{k'}e^{i(\varphi_{\mathbf{k}}-\varphi_{\mathbf{k}'})}+\sin^2\theta_k\sin^2\theta_{k'}e^{i(\varphi_{\mathbf{k}'}-\varphi_{\mathbf{k}})})(\langle c_{\mathbf{k}}^\dagger v_{\mathbf{k}}\rangle v_{\mathbf{k}'}^\dagger c_{\mathbf{k}'}+c_{\mathbf{k}}^\dagger v_{\mathbf{k}}\langle v_{\mathbf{k}'}^\dagger c_{\mathbf{k}'}\rangle-\langle c_{\mathbf{k}}^\dagger v_{\mathbf{k}}\rangle\langle v_{\mathbf{k}'}^\dagger c_{\mathbf{k}'}\rangle). \tag{G6}$$

Let $\Delta_{\mathbf{k}}=\sum_{\mathbf{k}'}V_{\mathbf{k},\mathbf{k}'}\langle v_{\mathbf{k}'}^\dagger c_{\mathbf{k}'}\rangle$, where $V_{\mathbf{k},\mathbf{k}'}=\frac{2\pi e^2}{\epsilon(\kappa+|\mathbf{k}-\mathbf{k}'|)}(\cos^2\theta_k\cos^2\theta_{k'}e^{i(\varphi_{\mathbf{k}}-\varphi_{\mathbf{k}'})}+\sin^2\theta_k\sin^2\theta_{k'}e^{i(\varphi_{\mathbf{k}'}-\varphi_{\mathbf{k}})})$. The total mean-field Hamiltonian is

$$\bar{H}=\sum_{\mathbf{k}}\Psi_{\mathbf{k}}^\dagger\begin{pmatrix}\sqrt{k^2+m^2} & -\Delta_k\\ -\Delta_k^\dagger & -\sqrt{k^2+m^2}\end{pmatrix}\Psi_{\mathbf{k}}+\sum_{\mathbf{k},\mathbf{k}'}V_{\mathbf{k},\mathbf{k}'}\langle c_{\mathbf{k}}^\dagger v_{\mathbf{k}}\rangle\langle v_{\mathbf{k}'}^\dagger c_{\mathbf{k}'}\rangle. \tag{G7}$$

The ground state energy is

$$E_g=-\sum_{\mathbf{k}}\sqrt{k^2+m^2+|\Delta_{\mathbf{k}}|^2}+\sum_{\mathbf{k}}\Delta_{\mathbf{k}}\langle c_{\mathbf{k}}^\dagger v_{\mathbf{k}}\rangle. \tag{G8}$$

According to

$$\frac{\delta E_g}{\delta\Delta_{\mathbf{k}}}=0, \tag{G9}$$

we obtain

$$\Delta_{\mathbf{k}}=\sum_{\mathbf{k}'}\frac{2\pi e^2}{\epsilon(\kappa+|\mathbf{k}-\mathbf{k}'|)}(\cos^2\theta_k\cos^2\theta_{k'}e^{i(\varphi_{\mathbf{k}}-\varphi_{\mathbf{k}'})}+\sin^2\theta_k\sin^2\theta_{k'}e^{i(\varphi_{\mathbf{k}'}-\varphi_{\mathbf{k}})})\frac{\Delta_{\mathbf{k}'}}{2\sqrt{k'^2+m^2+|\Delta_{\mathbf{k}'}|^2}}, \tag{G10}$$

where $|\mathbf{k}-\mathbf{k}'|=\sqrt{k^2+k'^2-2kk'\cos(\varphi_{\mathbf{k}}-\varphi_{\mathbf{k}'})}$. Subsequently, the summation in the above expression is replaced by an integral over momentum space

$$\Delta_{\mathbf{k}}=\frac{S}{(2\pi)^2}\int k'dk'd\varphi_{k'}\frac{2\pi e^2}{\epsilon(\kappa+|\mathbf{k}-\mathbf{k}'|)}(\cos^2\theta_k\cos^2\theta_{k'}e^{i(\varphi_{\mathbf{k}}-\varphi_{\mathbf{k}'})}+\sin^2\theta_k\sin^2\theta_{k'}e^{i(\varphi_{\mathbf{k}'}-\varphi_{\mathbf{k}})})\frac{\Delta_{\mathbf{k}'}}{2\sqrt{k'^2+m^2+|\Delta_{\mathbf{k}'}|^2}}. \tag{G11}$$

We choose a momentum cutoff $\Lambda$ then we can get

$$\begin{aligned}\frac{\Delta_{\mathbf{k}}}{\Lambda}=&\frac{S}{(2\pi)^2}\int_0^1\left(\frac{k'}{\Lambda}\right)d\left(\frac{k'}{\Lambda}\right)\int_0^{2\pi}d\varphi_{\mathbf{k}'}\frac{2\pi e^2}{\epsilon\left(\frac{\kappa}{\Lambda}+\sqrt{\left(\frac{k}{\Lambda}\right)^2+\left(\frac{k'}{\Lambda}\right)^2-2\frac{k}{\Lambda}\cdot\frac{k'}{\Lambda}\cdot\cos(\varphi_{\mathbf{k}}-\varphi_{\mathbf{k}'})}\right)}\\&\times\left(\cos^2\theta_{\frac{k}{\Lambda}}\cos^2\theta_{\frac{k'}{\Lambda}}e^{i(\varphi_{\mathbf{k}}-\varphi_{\mathbf{k}'})}+\sin^2\theta_{\frac{k}{\Lambda}}\sin^2\theta_{\frac{k'}{\Lambda}}e^{i(\varphi_{\mathbf{k}'}-\varphi_{\mathbf{k}})}\right)\frac{\frac{\Delta_{\mathbf{k}'}}{\Lambda}}{2\sqrt{\left(\frac{k'}{\Lambda}\right)^2+\left(\frac{m}{\Lambda}\right)^2+\left|\frac{\Delta_{\mathbf{k}'}}{\Lambda}\right|^2}}.\end{aligned} \tag{G12}$$

Then the condensed energy can be calculated by

$$E_c=-\sum_{\mathbf{k}}(\sqrt{k^2+m^2+|\Delta_{\mathbf{k}}|^2}-\sqrt{k^2+m^2})+\sum_{\mathbf{k}}\frac{|\Delta_{\mathbf{k}}|^2}{2\sqrt{k^2+m^2+|\Delta|^2}}, \tag{G13}$$

which can be written in the form of integration:

$$\begin{aligned}\frac{E_cU_0}{\Lambda^2}=&-\frac{S\Lambda U_0}{(2\pi)^2}\int_0^1\frac{k}{\Lambda}d\left(\frac{k}{\Lambda}\right)d\varphi_{\mathbf{k}}\left(\sqrt{\left(\frac{k}{\Lambda}\right)^2+\left(\frac{m}{\Lambda}\right)^2+\left(\frac{|\Delta_{\mathbf{k}}|}{\Lambda}\right)^2}\right.\\&\left.-\sqrt{\left(\frac{k}{\Lambda}\right)^2+\left(\frac{m}{\Lambda}\right)^2}\right)+\frac{S\Lambda U_0}{(2\pi)^2}\int_0^1\frac{k}{\Lambda}d\left(\frac{k}{\Lambda}\right)d\varphi_{\mathbf{k}}\frac{|\frac{\Delta_{\mathbf{k}}}{\Lambda}|^2}{2\sqrt{\left(\frac{k}{\Lambda}\right)^2+\left(\frac{m}{\Lambda}\right)^2+\left(\frac{|\Delta_{\mathbf{k}}|}{\Lambda}\right)^2}}.\end{aligned} \tag{G14}$$

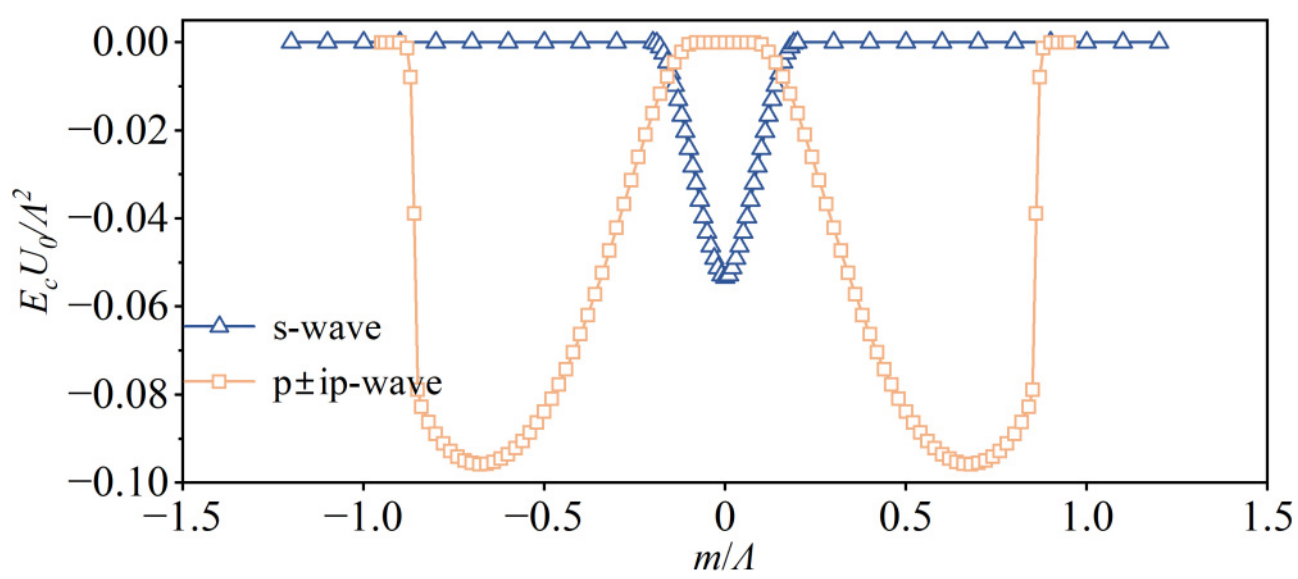

FIG. 7. The condensation energy of different phases with varying the renormalized gap $m$ for long-range interaction.

We take the parameter $\Lambda/\kappa = 0.1$ in this situation. As a result, for long-range interaction, we can plot the condensation energy of the $s$-wave and $p + ip$-wave excitonic insulator as shown in Fig. 7. This diagram is similar to Fig. 2, which indicates that the phase transition from the $s$-wave EI to the $p + ip$-wave EI exists. Similar order parameters give rise to similar ground state wave functions, so the typical long-range interactions characterizing these systems do not affect the quantum geometry discussed here.